\documentclass[preprint,3p,times,onecolumn]{elsarticle}
\usepackage{epstopdf}
\usepackage{graphicx}
\usepackage{amssymb}
\usepackage{tabularx}
\usepackage{hyperref}
\hypersetup{hidelinks,colorlinks=false,breaklinks=true,urlcolor= blue}
\usepackage{subcaption} 
\usepackage{amsmath}
\usepackage{pifont}
\newcommand{\cmark}{\text{\ding{51}}}
\newcommand{\xmark}{\text{\ding{55}}}
\usepackage{supertabular}
\usepackage{longtable}
\usepackage{float}

\usepackage{flushend}
\begin{document}

\begin{frontmatter}

\author[ADRESS]{Jonathan Dumas\corref{JDUMAS}}
\author[ADRESS]{Bertrand Corn\'elusse}
\address[ADRESS]{University of Li\`ege, Department of computer science and electrical engineering, Belgium}
\cortext[JDUMAS]{jdumas@uliege.be}

\title{Classification of load forecasting studies by forecasting problem to select load forecasting techniques and methodologies}

\begin{abstract}
	
The key contribution of this paper is to propose a classification into two dimensions of the load forecasting studies to decide which forecasting tools to use in which case. This classification aims to provide a synthetic view of the relevant forecasting techniques and methodologies by forecasting problem. In addition, the key principles of the main techniques and methodologies used are summarized along with the reviews of these papers.

The classification process relies on two couples of parameters that define a forecasting problem. Each article is classified with key information about the dataset used and the forecasting tools implemented: the forecasting techniques (probabilistic or deterministic) and methodologies, the data cleansing techniques, and the error metrics. 

The process to select the articles reviewed in this paper was conducted into two steps. First, a set of load forecasting studies was built based on relevant load forecasting reviews and forecasting competitions. The second step consisted in selecting the most relevant studies of this set based on the following criteria: the quality of the description of the forecasting techniques and methodologies implemented, the description of the results, and the contributions.

This paper can be read in two passes. The first one by identifying the forecasting problem of interest to select the corresponding class into one of the four classification tables. Each one references all the articles classified across a forecasting horizon. They provide a synthetic view of the forecasting tools used by articles addressing similar forecasting problems. Then, a second level composed of four Tables summarizes key information about the forecasting tools and the results of these studies. The second pass consists in reading the key principles of the main techniques and methodologies of interest and the reviews of the articles.

\noindent
\textbf{Word count: $\mathbf{20,000}$.}

\section*{Highlights}
\begin{itemize}
	\itemsep0em 
	\item A two-dimensional classification of load forecasting studies taking into account the system size
	\item A classification and a review with key information about the forecasting tools implemented and the dataset used  
	\item A way to select the relevant forecasting tools based on the definition of a forecasting problem
	\item A description of the main forecasting tools: the forecasting techniques and methodologies, the cleansing data techniques and the error metrics
\end{itemize}
\end{abstract}

\begin{keyword}
load forecasting \sep classification \sep forecasting techniques \sep forecasting methodologies \sep data cleansing techniques
\end{keyword}

\end{frontmatter}

\section{Introduction}\label{introduction}

\begin{figure}[htb]
	\centering
	\includegraphics[width=0.5\linewidth]{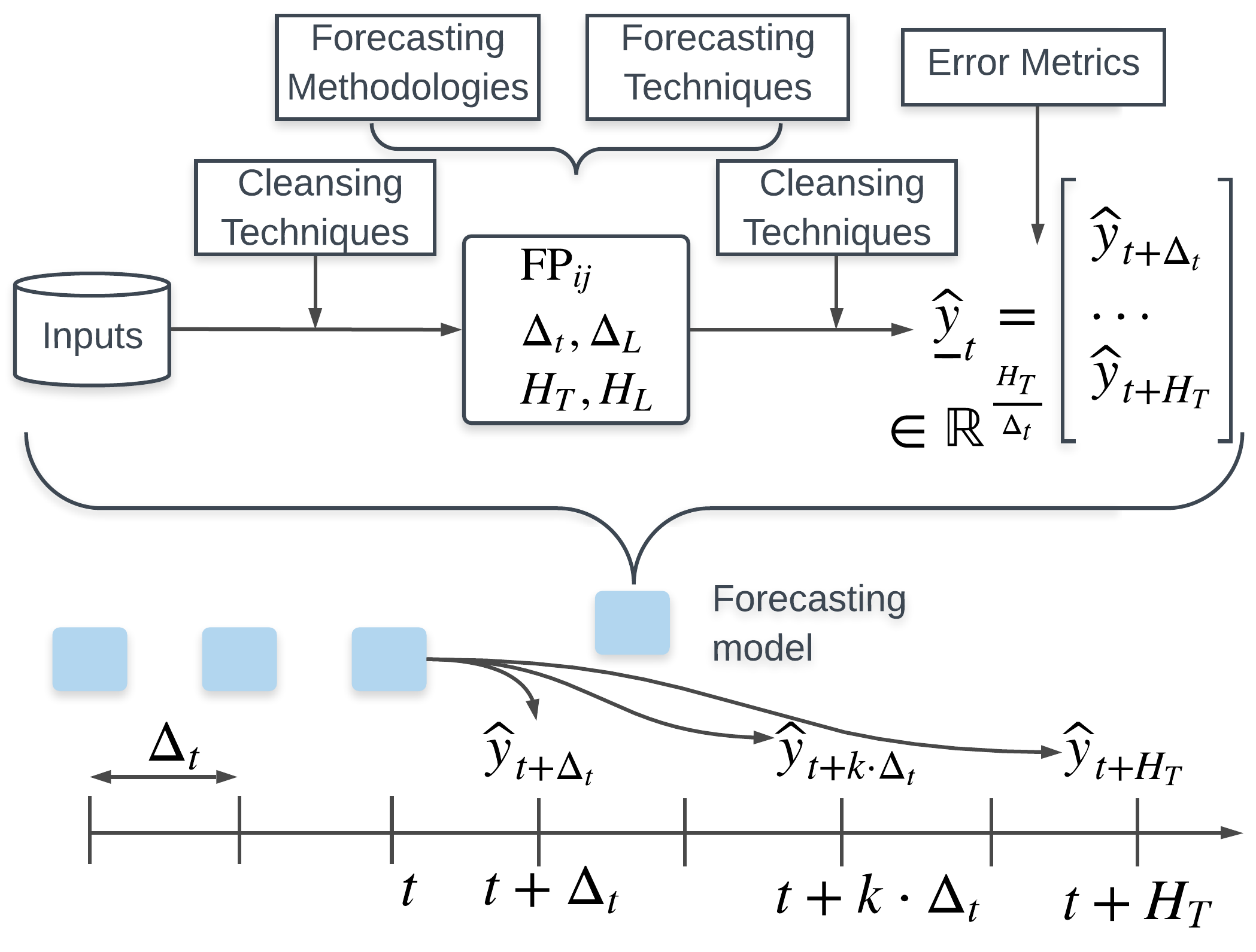}
	\caption{Forecasting model and process.}
	\label{fig:forecasting_model}
\end{figure}

Load forecasting arises in a wide range of applications such as providing demand response services for distribution or transmission system operators, bidding on the day ahead or future markets, power system operation and planning, or energy policies.
Each application is specific and requires a dedicated load forecasting model made of several modules, as depicted in Figure \ref{fig:forecasting_model}. The methodology to build a forecasting model is similar to the one required to solve a data science problem. The first step consists in understanding and formulating the problem: what is the application? What are the targets? This step should determine the forecasting horizon $H_T$, the temporal resolution $\Delta_t$, the system size $H_L$, and the spatial resolution $\Delta_L$ of the forecasting model. The second step consists in identifying, collecting, cleaning, and storing the data. The exploration and the analysis of the data are prerequisites for the next step to identify the relevant forecasting tools, the forecasting technique (FT), and the forecasting methodology (FM). The last step is the identification of the Error Metrics (EM) to assess the forecast vector $\widehat{\underline{y}}_t $. 

As mentioned above, the identification and selection of the relevant forecasting tools is problem dependent. Although the load forecasting literature is composed of several review papers, it is difficult to identify similar forecasting problems as there is no standard problem classification. However, most of the time the forecasting processes are classified based on the forecasting horizon, from very short-term to long-term, and by their deterministic or probabilistic nature.
\citet{tzafestas2001computational} review four classes of machine learning techniques for short-term load forecasting. 
\citet{alfares2002electric} review nine forecasting techniques and mentioned the horizon based classification but did not use it. 
\citet{Hong2016} review probabilistic forecasting across all forecasting horizons, and provide an overview of representative load forecasting techniques, methodologies and criteria to evaluate probabilistic forecasts. The spatial dimension is mentioned in the introduction with hierarchical load forecasting from the household level to the corporate level, across all forecasting horizons. However, it is considered as a separated topic and the spatial criterion is not combined with the horizon criterion.
\citet{van2017review} review the probabilistic load and PV forecasting techniques and performance metrics by using the forecasting horizon criterion. One of the objectives of this review is to find a common ground between probabilistic load and PV forecasting to address the problem of net demand forecasting.
\citet{deb2017review} review nine popular forecasting techniques and hybrid models, consisting in combination of two or more of these techniques, for load forecasting. A qualitative and quantitative comparative analysis of these techniques is conducted. The comparison of the pros and cons for each technique and the summary of the novelties brought by the hybrid models are useful to assess their potential. However, there is no classification across temporal or spatial dimensions.
These reviews provide useful references of load forecasting studies tackling deterministic or probabilistic forecasting problems along with information about forecasting methodologies and performance metrics. However, the spatial dimension is not taken into account and there is not a synthetic view of the studies classified with the key forecasting tools.

\begin{figure}[htb]
	\centering
	\includegraphics[width=0.5\linewidth]{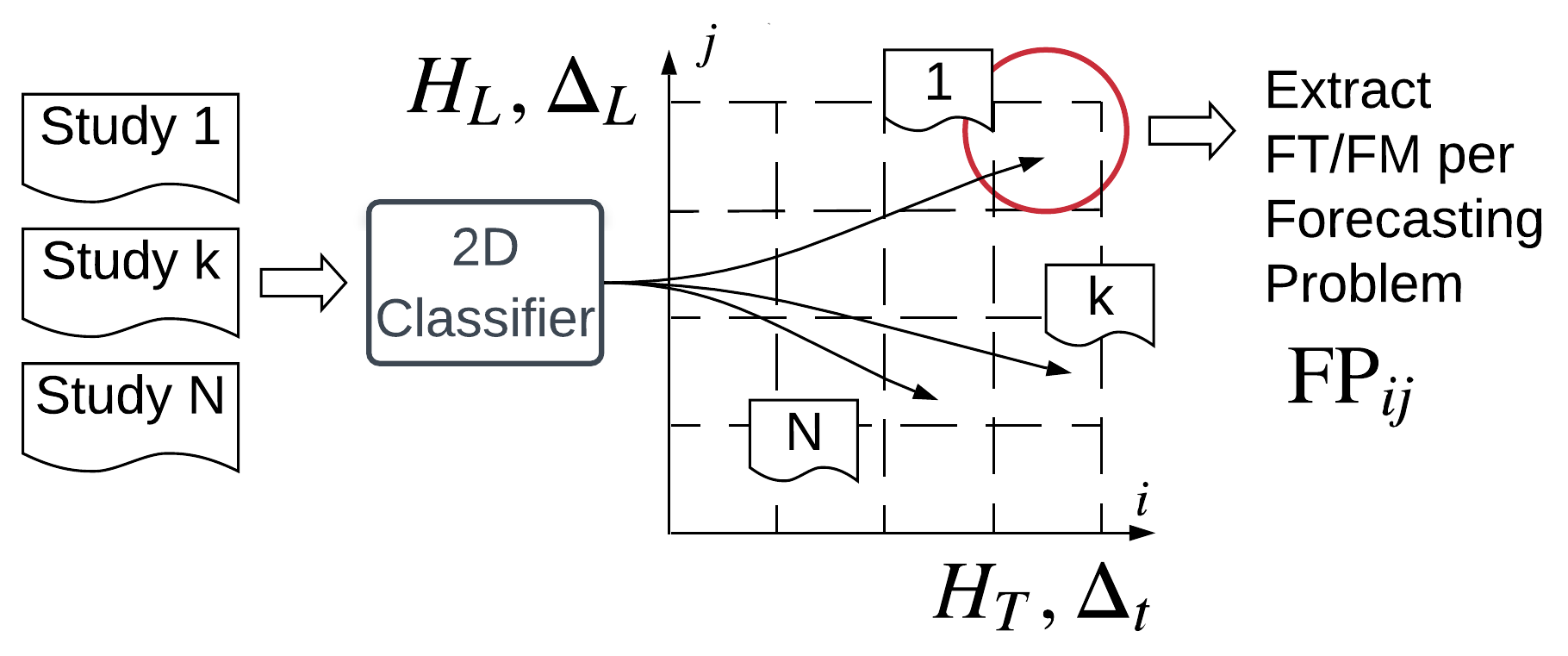}
	\caption{Two-dimensional classifier.}
	\label{fig:2D_classifier}
\end{figure}

To overcome this difficulty, this review adds the spatial dimension to the classification process to take into account the system size and describes, for each selected study, the forecasting techniques and methodologies implemented. The classification process is based on the definition of a forecasting problem with two couples of parameters. A temporal couple with the forecasting horizon and the resolution. A load couple with the system size and the load resolution. Each study reviewed is classified into a forecasting problem with key information about the forecasting tools implemented and the dataset used as shown on Figure \ref{fig:2D_classifier}.

\begin{figure}[htb]
	\centering
	\includegraphics[width=0.5\linewidth]{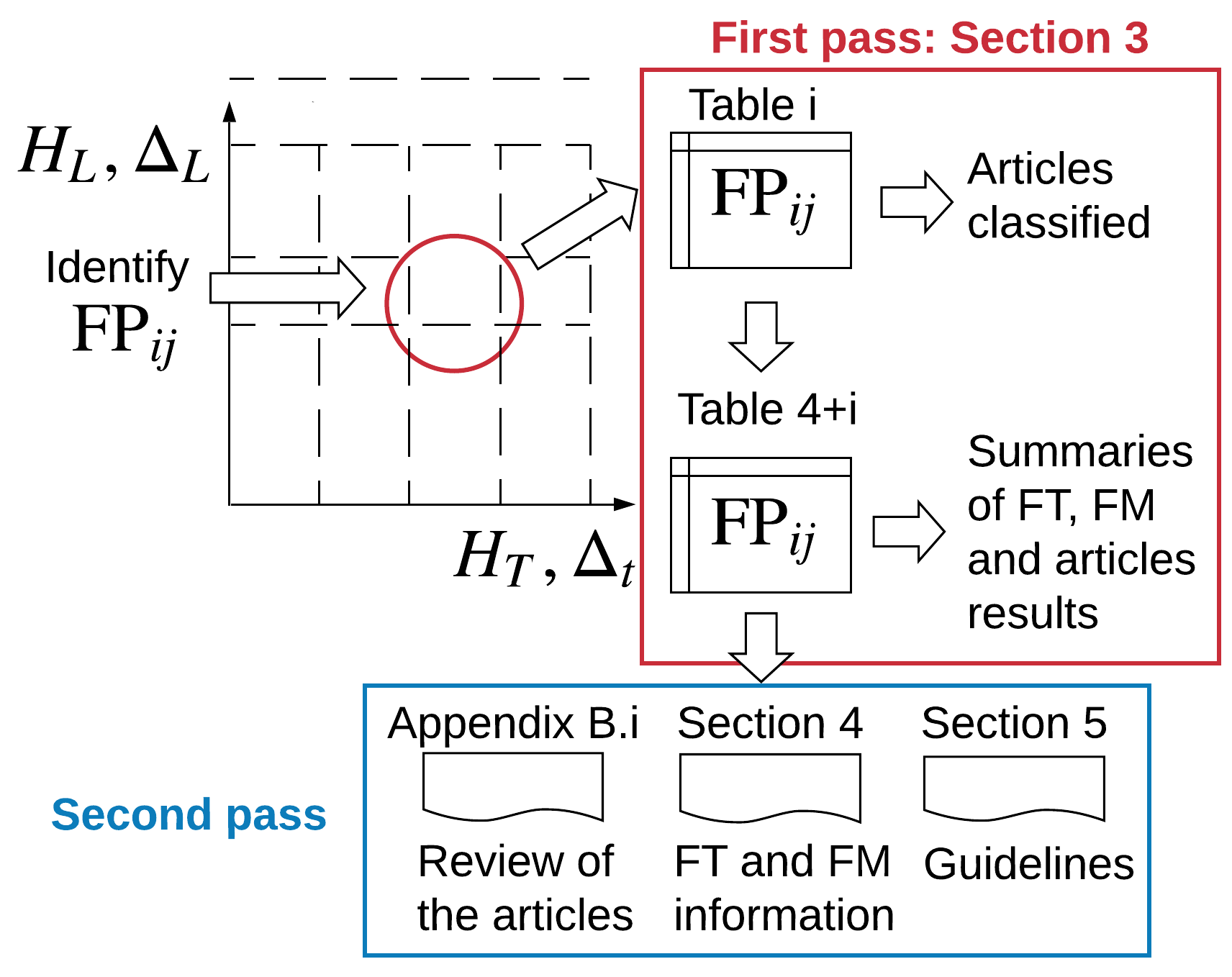}
	\caption{Methodology to use this review.}
	\label{fig:how_to_use_the_review}
\end{figure}

The methodology to use this review is described in Figure \ref{fig:how_to_use_the_review}. When the forecasting problem has been identified, it is possible to select the corresponding class $\text{FP}_{ij}, 1 \leq i,j \leq 4$ and to refer to the corresponding classification Table $i$. With $i=1,2,3,4$ corresponding to very short, short, medium, and long-term forecasting. Table $i$ references all the articles classified across the forecasting horizon class considered $i$ with the names of the forecasting techniques and methodologies used. It provides a synthetic view of the forecasting tools used by articles tackling similar forecasting problems. Then, Table $4+i$ summarizes key information about the forecasting tools and the results of these studies. Finally, Appendix B.i provides the review of these studies.

The more studies are classified the more forecasting techniques and methodologies are extracted by forecasting problem and the easier it should be to select the relevant ones. Indeed, more and more relevant techniques and methodologies should arise naturally from a given forecasting problem. 
The process to select the articles reviewed in this paper was conducted into two steps. First, a set of load forecasting studies was built based on relevant load forecasting reviews such as \citet{Hong2016, van2017review, deb2017review}, and forecasting competitions such as the Global Energy Forecasting Competitions 2012 and 2014 \cite{hong2014global, hong2016probabilistic}. These competitions are valuable sources to compare forecasting results from several combinations of forecasting techniques and methodologies as the forecasting problems and the datasets are the same for all the participants. 
The second step consisted in selecting the most relevant studies of this set based on the following criteria: the quality of the description of the forecasting techniques and methodologies implemented, the description of the results and the contributions. Figure \ref{fig:classification_overview} provides an overview of the articles classified by forecasting problem. However, almost half of the articles are classified within short and medium-term horizons for small load system. These articles are the top entries of the Global Energy Forecasting Competitions 2012 and 2014. Long-term horizon and medium load system classes have received less interest in the literature. This paper aims at providing a synthetic view of the relevant forecasting tools for a given forecasting problem, and the classification process was limited to keep the document readable. A possible extension is discussed in the conclusion.

This review has three main objectives: classifying the load forecasting studies of the literature into two dimensions leading to a set of sixteen classes; providing for each forecasting problem $\text{FP}_{ij}$ a synthetic view of the relevant forecasting tools (forecasting techniques and methodologies); ease the selection of the relevant forecasting tools to use in which case.

The paper structure is in line with the methodology depicted in Figure \ref{fig:how_to_use_the_review}.
Section \ref{sec:lf_problem} defines the time and load couples' classification parameters to build the classifier. Section \ref{sec:lf_classification} classifies the reviewed articles by forecasting problem. Tables \ref{tab:VSTLF}, \ref{tab:STLF}, \ref{tab:MTLF} and \ref{tab:LTLF} provide a synthetic view of the forecasting tools implemented and the datasets used for each article classified. Tables \ref{tab:VSTLF_key_information}, \ref{tab:STLF_key_information_1}, \ref{tab:STLF_key_information_2}, \ref{tab:MTLF_key_information_1}, \ref{tab:MTLF_key_information_2}, and \ref{tab:LTLF_key_information} summarize key information about the forecasting tools and the results. Section \ref{sec:lf_tools} focuses on the forecasting tools. The forecasting techniques and the methodologies implemented in the articles reviewed are succinctly summarized and references are given for further details. Finally, Section \ref{sec:guidelines} proposes general guidelines. Notations are provided in \ref{annex:notation}. The reviews of the studies are in \ref{annex:Review}. The datasets, data cleansing techniques, and error metrics used are referenced in \ref{annex:Datasets}, \ref{annex:DCT}, and \ref{annex:EM}.

\section{Load forecasting problem definition}\label{sec:lf_problem}

The following subsections define the parameters of a load forecasting problem.

\subsection{Temporal forecasting horizon $H_{T}$}

\citet{Hong2016} proposed a classification based on the forecast horizon into four categories: very short-term load forecasting (VSTLF), short-term load forecasting (STLF), medium-term load forecasting (MTLF), and long-term load forecasting (LTLF). The cut-off horizons are one day, two weeks, and three years respectively. The cut-off motivations are provided by \citet{hong2010short} and based on weather, economics and land use information.

The one-day cut-off between very short and short-term load forecasting is related to the temperature sensibility. In very short-term load forecasting, the temperature is relatively stable as the horizon is from a few hours to maximum one day. Thus, load is assumed to be weakly impacted by the temperature and can be forecasted only by its past values. However, depending on the system temperature load sensitivity this cut-off can be discussed. The two weeks cut-off between short and medium-term load forecasting is related to the temperature forecasts unreliability above this horizon. In short-term load forecasting, load is assumed to be affected significantly by the temperature. However, as in very short-term load forecasting, both economics and land use information are relatively stable within this horizon and are not necessarily required. The three years cut-off between medium- and long-term load forecasting is related to the economics forecasts unreliability above this horizon. In medium-term load forecasting, economics is required and predictable, temperature is simulated as no forecasts are reliable and land use is optional as it is stable within this horizon. In long-term load forecasting, temperature and economics are simulated and land use forecasts are used. Above a horizon of five years land use forecasts become unreliable and are simulated.

A cut-off is always questionable as it depends on the system considered and the forecast application. For large systems such as states or regions, forecasts with horizons up to a few years depend on macroeconomic indicators (such as gross domestic product). However, smaller systems, such as microgrids or small industries, depend on other economic indicators with shorter horizons. The choice of a one-year cut-off between medium and long-term load forecasting is done to find a trade-off between large and small systems. For all the other categories, the cut-offs are the same as those provided in \citet{hong2010short}.

\subsection{Temporal forecasting resolution $\Delta_t$} 

The temporal forecasting resolution is the time interval between each point of a forecast. It should not be confused with the updating resolution that is the time interval at which the forecasts are being updated. A very short-term load forecasting model can be updated each hour whereas a long-term load forecasting model is more likely to be updated monthly or yearly. The updating resolution belongs to the training methodology and is not a classification criterion. 

The temporal forecasting resolution ranges from a few minutes to years depending on the forecasting horizon. Usually the smaller horizon, the smaller the resolution is. Typical values are: from minutes to hours for very short-term load forecasting and short-term load forecasting, from hours to days for medium-term load forecasting and from days to years for long-term load forecasting.

\subsection{Load system size $H_{L}$} 

The load system size is related to the load capacity of the system considered. It is possible to classify a load system into four categories: very small load system (VSLS), small load system (SLS), medium load system (MLS) and large load system (LLS). Very small load systems are residential areas, small industrials or microgrids with load values from a few kW to MW, small load systems are thousands of residential areas, large industrials or microgrids from a few MW to GW, medium load systems are regional or small state grids from a few GW to ten GW, and large load systems are large state to continental grids from ten GW to a hundred GW.

\subsection{Load forecasting resolution $\Delta_L$}

The load forecasting resolution is the resolution of the forecasts. Usually, the smaller the load system size, the smaller the load resolution is. Typical values are from W to kW for very small load system, kW to MW for small load system and medium load system, MW to GW for large load system.

\subsection{A two-dimensional classification process}

\begin{figure}[tb]
	\begin{subfigure}[b]{0.4\textwidth}
	\includegraphics[width=8cm,height=6cm]{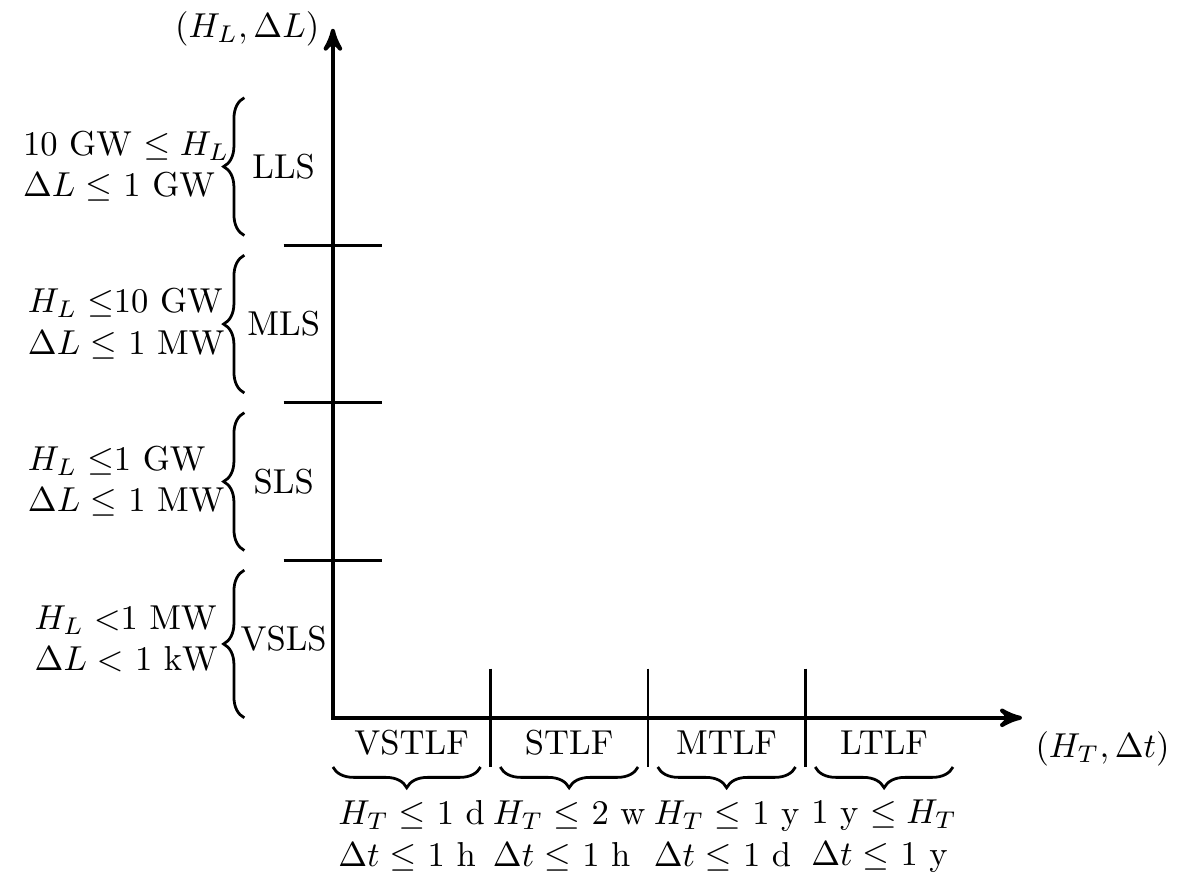}
	\caption{Classes of the two-dimensional classifier.}
	\label{classes}
	\end{subfigure}
	\begin{subfigure}[b]{0.4\textwidth}
	\includegraphics[width=8cm,height=6cm]{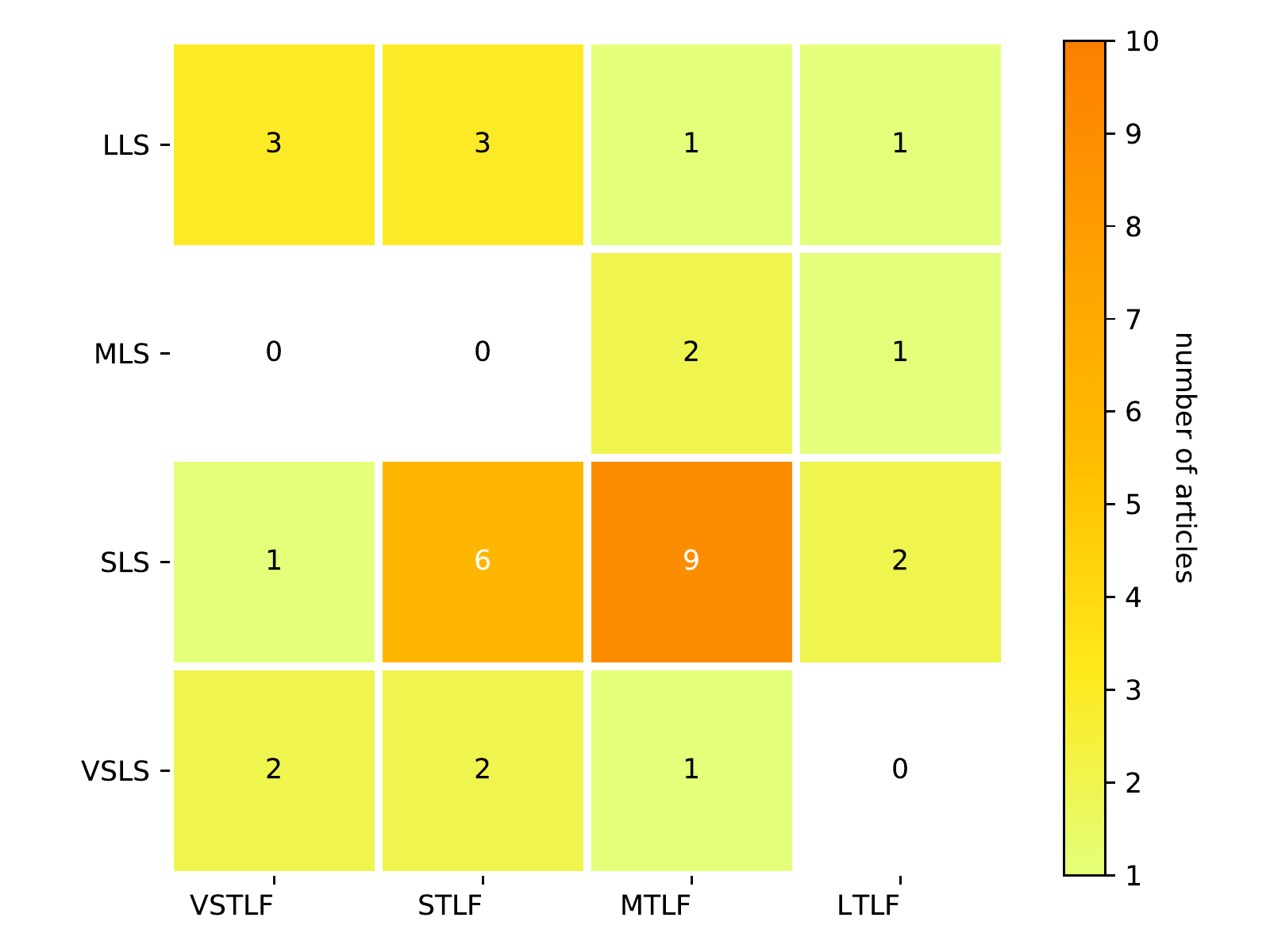}
	\caption{Articles classified by forecasting problem.}
	\label{fig:classification_overview}
	\end{subfigure}
	\caption{Classification of load forecasting studies.}
\end{figure}

The four parameters $H_{T}$, $\Delta_t$, $H_{L}$ and $\Delta_L$ define a four dimensional classifier. However, as $\Delta_t$ is related to $H_{T}$ and $\Delta_L$ to $H_{L}$, it is possible to define a two-dimensional classifier by considering the temporal and load couples $(H_{T}, \Delta_t)$ and $(H_{L}, \Delta_L)$ denoted by the integers $i$ and $j$, respectively. Figure \ref{classes} shows the sixteen forecasting problems defined by this classifier. With $i=1,2,3,4$ corresponding to very short, short, medium, and long-term forecasting. And with $j=1,2,3,4$ corresponding to very small, small, medium, and large systems.

\newpage

\section{Two-dimensional load forecasting classification}\label{sec:lf_classification}

Figure \ref{fig:classification_overview} provides a synthetic view of the classification by forecasting problem of the reviewed articles, according to the selection process explained in the introduction\footnote{These articles do not give an exhaustive overview of the load forecasting literature. It has to be pointed out that a paper can be classified in several classes. Indeed, it may investigate several forecasting horizons or system sizes. When it happens, there is only one review of the article in \ref{annex:Review}.}. Tables \ref{tab:VSTLF}, \ref{tab:STLF}, \ref{tab:MTLF}, and \ref{tab:LTLF} correspond to very short, short, medium, and long-term forecasting, respectively. They provide a synthetic view of the forecasting tools and datasets used. Within each table the studies are classified from  top to  bottom by increasing the system size. The information provided are the references, the classes $[i,j]$, the values of $\Delta_L$, $\Delta_t$, $H_{T}$ and $H_{L}$, the names of the forecasting techniques and methodologies, data cleansing techniques, error metrics, and the datasets used. Tables \ref{tab:VSTLF_key_information}, \ref{tab:STLF_key_information_1}, \ref{tab:STLF_key_information_2}, \ref{tab:MTLF_key_information_1}, \ref{tab:MTLF_key_information_2}, and \ref{tab:LTLF_key_information} summarize key information about the forecasting tools and the results. \ref{annex:Review_VSTLF} to \ref{annex:Review_LTLF} provide the review of these papers. Tables \ref{tab:LFdatareview1}, \ref{tab:LFdatareview2}, \ref{tab:DCT}, \ref{tab:EMdeterministic}, and \ref{tab:EMproba} in \ref{annex:Datasets}, \ref{annex:DCT},  and \ref{annex:EM} provide key information about the datasets, data cleansing techniques and error metrics used.


\begin{table*}[!htb]
	\renewcommand\thetable{1}
	\renewcommand{\arraystretch}{1.2}
	\caption{Very Short-Term Load Forecasting classification. \label{tab:VSTLF}} 
	\begin{tabular}{m{1.5cm}m{0.6cm}||m{0.5cm}m{0.7cm}m{0.7cm}m{0.5cm}|m{4cm}m{2.5cm}m{0.2cm}m{0.2cm}m{0.5cm}}
		Author & Class & $\Delta_L$ & $\Delta_t$ & $H_{T}$ & $H_{L}$ & Techniques & Methodologies   & PF  & DC  & Data  \\ \hline \hline
		\citet{VanderMeerShepero2018} & $[1,1]$  & - & 30 m & 30 m & 10 kW & ARIMA, Gaussian process  & VS, TM & \cmark & \cmark &  \hyperref[D15]{D15}  \\
		\citet{Chae2016} &  $[1,1]$ & kW & 15 m & 1 d & 800 kW & Neural network  & VS, TM & \xmark & \cmark &  site's data \\
		\citet{Wang2016} & $[1,2]$ & kW & 1 h & 1 d & 2 GW & Linear regression  & VS, VWS, THF, LHF, TM & \xmark & \xmark &   \hyperref[D1]{D1}  \\
		\citet{taylor2007short} & $[1,4]$ & MW & 30 m - 1 h & 1 h - 24 h & 50 GW & ARMA, AR, Exponential smoothing, Principal component analysis  & VS, TM & \xmark & \cmark &   \hyperref[D14]{D14}  \\
		\citet{Taylor2008} & $[1,4]$ & MW & 1 m &  1 m - 24 h  & 50 GW & ARMA, Neural network, Exponential smoothing, weather-based approach  & VS, FSAC, SD & \xmark & \cmark &   \hyperref[D13]{D13}  \\
		\citet{Taylor2010} & $[1,4]$ & MW & 30 m & 1 h - 24 h & 100 GW & ARMA, Neural network, Exponential smoothing   & VS, THF, FSAC& \xmark & \cmark &  \hyperref[D14]{D14} 
	\end{tabular}
\end{table*}

\newpage

\begin{table*}[!htb]
	\renewcommand\thetable{2}
	\renewcommand{\arraystretch}{1.2}
	\caption{Short-Term Load Forecasting classification. \label{tab:STLF}} 
	\begin{tabular}{m{1.5cm}m{0.6cm}||m{0.5cm}m{0.7cm}m{0.7cm}m{0.5cm}|m{4cm}m{2.5cm}m{0.2cm}m{0.2cm}m{0.5cm}}
		Author & Class & $\Delta_L$ & $\Delta_t$ & $H_{T}$ & $H_{L}$ & Techniques & Methodologies   & PF  & DC  & Data  \\ \hline \hline
		\citet{Shepero2018} & $[2,1]$ & - & 30 m & 1 w & 10 kW & Gaussian process & VS & \cmark & \cmark & \hyperref[D15]{D15} \\
		\citet{Ahmad2018} & $[2,1]$ & kW & 5 m & 1 w - 1 M & 500 kW & Decision tree, $k$-nearest neighbor, linear regression  & VS & \xmark & \cmark & site's data \\
		\citet{Goude2014} & $[2,2]$ & - & 10 m &1 d &  50 MW & General additive models  & VS, LHF, THF, MWS & \xmark & \cmark  &  \hyperref[D3]{D3} \\
		\citet{Charlton2014} & $[2,2]$ & kW & 1 h &1 w &  2 MW & Linear regression  & VS, LHF, THF, MWS, LA, FSAC & \xmark & \cmark  &  \hyperref[D1]{D1} \\
		\citet{Lloyd2014} & $[2,2]$ & kW & 1 h &1 w &  2 MW & Gradient boosting, Gaussian process, linear regression  & VS, FWC, LHF & \xmark & \cmark  &  \hyperref[D1]{D1} \\
		\citet{Nedellec2014} & $[2,2]$ & kW & 1 h &1 w &  2 MW & General additive models, Random forest  & VS, LHF, THF, MWS & \xmark & \xmark  &  \hyperref[D1]{D1} \\
		\citet{bentaieb2013} & $[2,2]$ & kW & 1 h &1 w &  2 MW & Gradient boosting, general additive models  & LHF, THF, VS, MWS & \xmark & \cmark  &  \hyperref[D1]{D1} \\
		\citet{hong2010short} & $[2,2]$ & kW & 1 h &1 h - 1 y &  1 GW & Linear and Fuzzy interaction regressions, Neural network & VS, THF, TM & \xmark & \xmark  &  \hyperref[D2]{D2} \\
		\citet{HongWang2014} & $[2,4]$ & MW & 1 h &1 d &  25 GW & Linear and Fuzzy interaction regressions & VS, TM & \xmark & \xmark  &  \hyperref[D10]{D10} \\
		\citet{Qahtani2013} & $[2,4]$ & MW & 1 h & 1 d&  50 GW & $k$-nearest neighbor & VS, TM & \xmark & \xmark  &  \hyperref[D13]{D13} \\
		\citet{Lopez2018} & $[2,4]$ & MW & 1 h &9 d &  40 GW & Neural network, AR & VS, LHF, MWS, FWC, TM & \cmark & \cmark  &  \hyperref[D5]{D5}
	\end{tabular}
\end{table*}

\newpage

\begin{table*}[!htbp]
	\renewcommand\thetable{3}
	\renewcommand{\arraystretch}{1.2}	
	\caption{Medium-Term Load Forecasting classification. \label{tab:MTLF}} 
	\begin{tabular}{m{1.5cm}m{0.6cm}||m{0.5cm}m{0.7cm}m{0.7cm}m{0.5cm}|m{4cm}m{2.5cm}m{0.2cm}m{0.2cm}m{0.5cm}}
		Author & Class & $\Delta_L$ & $\Delta_t$ & $H_{T}$ & $H_{L}$ & Techniques & Methodologies   & PF  & DC  & Data  \\ \hline \hline
		\citet{Xie2015LTLF} & $[3,1]$ & kW & 1 h & 9 M &  5 kW & Linear regression & VS, WSS & \xmark & \xmark  & \hyperref[D8]{D8} \\
		\citet{Xie2015} & $[3,2]$ & kWh & 1 d & 6 M & 4 MW & Linear regression, ARIMA, Neural network, Random forest  & FSAC, VS & \xmark & \xmark &   \hyperref[D4]{D4} \\
		\citet{Xie2015LTLF} & $[3,2]$ & MW & 1 h & 9 M & 19 MW & Linear regression  & VS, WSS, FC & \xmark & \xmark &   \hyperref[D8]{D8} \\
		\citet{Goude2014} & $[3,2]$ & - & 10 m & 1 y &  50 MW & General additive model & VS, LHF, THF, MWS  & \xmark & \cmark  & \hyperref[D3]{D3} \\
		\citet{Gaillard2016} & $[3,2]$ & 0.1 MW & 1 h & 1 M & 310 MW & General additive model, Quantile regression  & VS, THF, VWS, DS & \cmark & \xmark &   \hyperref[D6]{D6} \\
		\citet{Dordonnat2016} & $[3,2]$ & 0.1 MW & 1 h & 1 M & 310 MW & General additive model  & VS, VWS & \cmark & \xmark &   \hyperref[D6]{D6} \\
		\citet{xie2016} & $[3,2]$ & 0.1 MW & 1 h & 1 M & 310 MW & Linear regression, Unobserved components model, Exponential smoothing, Neural network, ARIMA  & FSAC, VWS, VS, RS & \cmark & \xmark &   \hyperref[D6]{D6} \\
		\citet{Haben2016} & $[3,2]$ & 0.1 MW & 1 h & 1 M & 310 MW & Kernel density estimation, Quantile regression  & THF, VWS & \cmark & \xmark &   \hyperref[D6]{D6} \\
		\citet{Chen2001} & $[3,2]$ & MW & 1 d &1 M &  800 MW & Support vector regression & VS, DS & \xmark & \xmark  &  \hyperref[D16]{D16} \\
		\citet{Wang2016} & $[3,3]$ & kW & 1 h & 1 y & 2 GW & Linear regression  & VS, VWS, LHF, TM & \xmark & \xmark &   \hyperref[D1]{D1} \\
		\citet{Ziel2016} & $[3,3]$ & MW & 1 h & 1 y & 3.3 GW & AR  & VS, VWS & \cmark & \xmark &   \hyperref[D6]{D6}, \hyperref[D7]{D7}\\
		\citet{Xie2017} & $[3,3]$ & MW & 1 h & 1 y & 5 GW & Linear regression, Neural network  &VS, RS & \cmark & \xmark &   \hyperref[D6]{D6}, \hyperref[D9]{D9}\\
		\citet{hyndman2015monash} & $[3,4]$ & - & - & 1 w - 1 y & 30 GW & General additive model  &THF, VS, VWS, DS & \cmark & \xmark &   \hyperref[D12]{D12}
	\end{tabular}
\end{table*}	

\newpage


\begin{table*}[!htbp]
	\renewcommand\thetable{4}
	\renewcommand{\arraystretch}{1.2}	
	\caption{Long-Term Load Forecasting classification. \label{tab:LTLF}} 
	\begin{tabular}{m{1.5cm}m{0.6cm}||m{0.5cm}m{0.7cm}m{0.7cm}m{0.5cm}|m{4cm}m{2.5cm}m{0.2cm}m{0.2cm}m{0.5cm}}
		Author & Class & $\Delta_L$ & $\Delta_t$ & $H_{T}$ & $H_{L}$ & Techniques & Methodologies   & PF  & DC  & Data  \\ \hline \hline
		\citet{Hong2008} & $[4,2]$  &  10 kW & 1 y & 20 y & 500 MW & S-curve fitting & LHF, Human-Machine Co-construct Intelligence & \xmark & \xmark & \hyperref[D17]{D17} \\
		\citet{HongXie2014} & $[4,3]$ & MW & 1 h - 1 y & 1 y & 5 GW & Linear regression  & VS & \cmark & \xmark &   \hyperref[D9]{D9}\\
		\citet{HongShahidehpour2015} & $[4,4]$ & MW & 1 h - 1 y & 1 y & 5 GW & Linear regression  & VS, LHF, WSS & \cmark & \xmark &   \hyperref[D9]{D9}, \hyperref[D10]{D10}, \hyperref[D11]{D11}
	\end{tabular}
\end{table*}
\newpage


\begin{table*}[!htb]
	\renewcommand\thetable{5}
	\renewcommand{\arraystretch}{1.2}
	\caption{Very Short-Term Load forecasting summary. \label{tab:VSTLF_key_information}} 
	\begin{tabular}{m{1.5cm}m{0.6cm}||m{13.5cm}}
		Author & Class & Forecasting tools and results \\ \hline \hline
		\citet{VanderMeerShepero2018} & $[1,1]$
		& Gaussian process is implemented and ARIMA used as benchmark. The dynamic Gaussian process uses a sliding window training methodology with a linear combination of the squared exponential and Mat\'ern kernels. Deterministic and probabilistic metrics are computed using a blocked form of $k$-fold cross-validation procedure.
		The Gaussian process models were consistently outperformed by ARIMA by a factor in terms of MAPE and NRMSE for the demand, PV production, and net demand forecasting. However, ARIMA produced higher prediction intervals that are constant over time. In contrast, the Gaussian process have wide and narrow prediction intervals between periods of high and low uncertainty.  \\ \hline
		\citet{Chae2016} & $[1,1]$ 
		& Random forest is used for feature extraction and a feed-forward artificial neural network was selected among nine machine learning techniques. A static, accumulative and sliding window training methodologies are implemented and deterministic metrics are computed using a blocked form of $k$-fold cross-validation procedure.
		The average value of the daily coefficient variance of the RMSE, that is a normalized RMSE, of the consumption is around 10 \% and the average APE of the daily peak demand is around 5 \%.   \\ \hline
		\citet{Wang2016}& $[1,2]$ 
		& A customized linear regression technique models the recency effect both at the aggregated and bottom levels and for each hour of the day.   
		A static and sliding window training methodologies are used to compute yearly and daily forecasts.
		The customized linear regression model significantly outperformed the benchmark with a MAPE reduced between 10 \% and 20 \% depending on the load level and the forecasting horizon.
		\\ \hline
		\citet{taylor2007short} & $[1,4]$ 
		& Several statistical univariate methods are implemented: double seasonal ARMA, periodic Autoregressive, double seasonal Holt-Winters exponential smoothing, intraday cycle exponential smoothing and principal component analysis.
		A static training methodology is used.
		The double seasonal Holt-Winters method performed the best followed by the principal component analysis and the seasonal ARMA. The MAPE are between 0.75 \% and 2.2 \%.
		\\ \hline 
		\citet{Taylor2008} & $[1,4]$
		& Several statistical univariate methods are implemented: double seasonal ARMA, double seasonal Holt-Winters exponential smoothing, double seasonal intraday cycle exponential smoothing, the weather-based approach used at National Grid, and a combination of the weather-based approach and the double seasonal exponential smoothing by taking the average. A static training methodology is used.
		For forecasting horizons between one and thirty minutes, the double seasonal exponential smoothing ranked first with MAPE values between 0.12 \% and 0.5 \%. For forecasting horizons between one minutes and one day,
		the combination approach performed the best with MAPE values between 0.15 \% and 1.2 \%.
		\\ \hline  
		\citet{Taylor2010}& $[1,4]$ 
		& Extension from double to triple statistical seasonal univariate methods are implemented along with a feed-forward neural network. A combination of the triple seasonal ARMA and Holt-Winter exponential smoothing approaches is also implemented. A static training methodology is used.
		The triple seasonal versions of all the methods were more accurate than the double ones with MAPE values from 0.4 \% to 1.75 \%. The combination approach performed the best results.
	\end{tabular}
\end{table*}

\newpage

\begin{table*}[!htb] 
	\renewcommand\thetable{6.1}
	\caption{Short-Term Load forecasting summary, part 1. \label{tab:STLF_key_information_1}} 
	\renewcommand{\arraystretch}{1.2}
	\begin{tabular}{m{1.5cm}m{0.6cm}||m{13.5cm}}
		Author & Class & Forecasting tools and results \\ \hline \hline
		\citet{Shepero2018} & $[2,1]$
		& A log-normal transformation of the data before applying Gaussian process is compared to the conventional Gaussian process approach. The training methodology is a blocked form of cross-validation.
		The log-normal transformation produced sharper forecasts than the conventional approach and the sharpness of the forecasts varied throughout the day. That is not the case without transformation. However, both approaches were not able to capture the sudden sharp increments of the load.  
		\\ \hline 
		\citet{Ahmad2018} & $[2,1]$
		& Compact decision trees, fit $k$-nearest neighbor, linear and stepwise linear regressions models are implemented. The inputs are climate variables, lags of load, and calendar variables. The training methodology is static.
		The $k$-nearest neighbor approach achieved the best MAPE with 0.076 \% followed by the decision trees for one week ahead forecasting. Inversely, the decision trees performed the best MAPE with 0.044 \% followed by the $k$-nearest neighbor for one month ahead forecasting.
		\\ \hline 
		\citet{Goude2014} & $[2,2]$
		& Refer to Table \ref{tab:MTLF_key_information_1} for description.
		\\ \hline 
		\citet{Charlton2014} & $[2,2]$
		& A linear regression model with temperature and calendar variables is implemented. Load and temporal hierarchical forecasting methodologies are used by computing a model for each zone, hours of the day, seasons of the year, weekdays, and weekend. A multiple weather station selection methodology computes for each zone a linear combination of the five best weather stations. Outliers removal is used.
		Ranked first at the Global Energy Forecasting Competition 2012.
		\\ \hline 
		\citet{Lloyd2014} & $[2,2]$
		& The final model is a weighted average of gradient boosting, Gaussian process, and linear regression models. There is a gradient boosting model per zone and the demand is modeled as a function of temperature and calendar variables. The Gaussian process kernels are the squared exponential and the periodic ones.
		Ranked second at the Global Energy Forecasting Competition 2012.   
		\\ \hline 
		\citet{Nedellec2014} & $[2,2]$
		& A multi-scale approach with three models that correspond to the long, medium, and short terms. General additive models are used for long and medium terms and  random forest for short-term. The inputs are the monthly load and temperature, the day type, the time of the year, and a smoothed temperature. There is a model for each load zone and a medium-term model is fitted per instant of the day.
		Ranked third at the Global Energy Forecasting Competition 2012. The short-term model provides an average gain of 5 \% in terms of RMSE.  
	\end{tabular}
\end{table*}

\newpage
\begin{table*}[!htb] 
	\renewcommand\thetable{6.2}
	\caption{Short-Term Load forecasting summary, part 2. \label{tab:STLF_key_information_2}} 
	\renewcommand{\arraystretch}{1.2}
	\begin{tabular}{m{1.5cm}m{0.6cm}||m{13.5cm}}
		Author & Class & Forecasting tools and results \\ \hline \hline
		\citet{bentaieb2013} & $[2,2]$
		& A gradient boosting approach to estimate general additive models. There is a model per load zone and for each hour of the day.
		Ranked fourth at the Global Energy Forecasting Competition 2012.
		\\ \hline 
		\citet{hong2010short} & $[2,2]$
		&  A systematic approach to investigate short-term load forecasting is conducted. A methodology to select the relevant variables is exposed. Several linear and fuzzy regression models, and single-output feed-forward neural networks with consecutive refinements are implemented. A sliding window methodology is used. 
		The case study demonstrated that linear regression models can be more accurate than feed-forward neural networks and fuzzy regression models given the same amount of input information.
		The MAPE of the linear regression benchmark and its customized version are approximately 5 \% and 3 \% for horizons of one hour, one day, one week, two weeks, and one year. The MAPE of the feed-forward neural network benchmark and its  customized versions are approximately 6.5 \% and 4.5 \% for an horizon of one year.
		\\ \hline 
		\citet{HongWang2014} & $[2,4]$
		& A fuzzy interaction regression approach with three refined versions are compared to a linear regression model. A sliding window methodology is used.
		The most refined version of the fuzzy interaction regression models performed the best MAPE with 3.68 \%.
		\\ \hline 
		\citet{Qahtani2013} & $[2,4]$
		& A univariate and multivariate $k$-nearest neighbor approaches are implemented. The training methodology is static. The multivariate approach takes lags of load and calendar variables as inputs.
		The MAPE of the multivariate and univariate approaches are 1.81 \% and 2.38 \%.  
		\\ \hline 
		\citet{Lopez2018} & $[2,4]$
		& An on-line real-time hybrid load-forecasting model based on an Autoregressive model and neural networks is studied. The training methodology is static. The final model at the national level is a linear combination of four sub-models. The model takes calendar information such as special days and temperature lags as inputs.
		The RMSE and MAPE of the final model are 1.83 \% and 1.56 \%. 	
	\end{tabular}
\end{table*}

\newpage


\begin{table*}[!htb]
	\renewcommand\thetable{7.1}
	\renewcommand{\arraystretch}{1.2}
	\caption{Medium-Term Load forecasting summary, part 1. \label{tab:MTLF_key_information_1}} 
	\begin{tabular}{m{1.5cm}m{0.6cm}||m{13.5cm}}
		Author & Class & Forecasting tools and results \\ \hline \hline
		\citet{Xie2015LTLF} & $[3,1]$
		& The linear regression benchmark of \citet{hong2010short} is used as a starting model. Then, the variable selection framework of \citet{hong2010short} is applied to customized the model. The training methodology is static. The customized linear regression model forecasts the load per customer. The survival analysis model forecasts the tenured customers. Then, the final forecast is the multiplication of both previous forecasts. This methodology is validated through a field implementation at a fast growing U.S. retailer. 
		The MAPE of the customized linear regression model is 11.56 \%. The MAPE of the final forecast is around 10.5 \%.
		\\ \hline 
		\citet{Xie2015} & $[3,2]$
		& A combination of four forecasting techniques is implemented: linear regression, ARIMA, feed-forward neural network, and random forest. A forward and a backward selection strategies are used to select the relevant variables of the linear regression model. The training methodology is static.
		This methodology ranked third at NPower Forecasting Challenge 2015. The MAPE of the final model is 2.40 \% and better than any of the individual model. The best linear regression model performed  2.47 \%, the neural network 3.27 \%, the random forest 3.87 \%, and ARIMA 8.23 \%.  
		\\ \hline
		\citet{Xie2015LTLF} & $[3,2]$
		& Refer to $[3,1]$ of this Table for description. 
		\\ \hline
		\citet{Goude2014} & $[3,2]$
		& A general additive model is implemented for one day ahead forecasting and two others for one year ahead forecasting. The training methodology is static. Load and temporal hierarchical forecasting methodologies are used. These models are able to forecast more than 2 000 electricity consumption series.
		The one day ahead model achieved a median MAPE of 5 \%. The one year ahead models achieved a median MAPE of 6 \% and 8 \%.   
		\\ \hline
		\citet{Gaillard2016} & $[3,2]$
		& A concatenation of a short and medium-term quantile generalized additive models is implemented. Temperature scenarios are used into the probabilistic forecasting load model. A weather station selection is used to compute an average temperature that is an input of the models. There is one model fitted per hour of the day.
		Ranked first at the Global Energy Forecasting Competition 2014. The PLF values are between 4 and 11.  
		\\ \hline
		\citet{Dordonnat2016} & $[3,2]$
		& A temperature based deterministic load model with a generalized additive model uses temperature sample paths to produce load paths. The final probabilistic forecast is derived from these final paths. A weather station selection is used to compute an average temperature as input of the model. 
		Ranked second at the Global Energy Forecasting Competition 2014. The MAPE values of the deterministic model are between 8.74 \% and 11.83 \%. The PLF of the probabilistic models are between 7.37 to 8.37.
	\end{tabular}
\end{table*}

\newpage

\begin{table*}[!htb]
	\renewcommand\thetable{7.2}
	\renewcommand{\arraystretch}{1.2}
	\caption{Medium-Term Load forecasting summary, part 2. \label{tab:MTLF_key_information_2}} 
	\begin{tabular}{m{1.5cm}m{0.6cm}||m{13.5cm}}
		Author & Class & Forecasting tools and results \\ \hline \hline
		\citet{xie2016} & $[3,2]$
		& An integrated solution with pre-processing, forecasting, and post-processing is proposed. The pre-processing is data cleansing and temperature weather station selection. The forecasting  part is achieved with a linear regression deterministic model combined with residual forecasts of four techniques: exponential smoothing, ARIMA, feed-forward neural network, and unobserved components model. Then, temperature scenarios are used to generate probabilistic forecasts. Finally, the post-processing uses residual simulation to improve the probabilistic forecasts.
		Ranked third at the Global Energy Forecasting Competition 2014. The residual simulation of the post-processing step helped to improve the forecasts. The PLF of the submitted model are between 3.360 and 11.867.
		\\ \hline  
		\citet{Haben2016} & $[3,2]$
		& A hybrid model of kernel density estimation and quantile regression is implemented. It relies on a combination of five models each one forecasting on a dedicated time period. 
		Ranked fourth at the Global Energy Forecasting Competition 2014.   
		\\ \hline
		\citet{Chen2001} & $[3,2]$
		& A support vector regression model using data segmentation is implemented. 
		It demonstrates that a conservative approach using only available correct information is recommended in this case due to the difficulty to provide accurate temperature forecast. Ranked first at the EUNITE Competition 2001. 
		\\ \hline 
		\citet{Wang2016} & $[3,3]$
		& Refer to Table \ref{tab:VSTLF_key_information} for description.
		\\ \hline 
		\citet{Ziel2016} & $[3,3]$
		& A methodology based on Least Absolute Shrinkage and Selection Operator (LASSO) and bivariate time-varying threshold Autoregressive model. The LASSO has the properties of automatically selecting the relevant variables of the model. Then, a residual-based bootstrap is used to simulate future scenario sample paths to generate probabilistic forecasts.
		Ranked second at the extended version of the Global Energy Forecasting Competition 2014. The PLF scores are 7.44 and 54.69 for both competitions.   
		\\ \hline
		\citet{Xie2017} & $[3,3]$
		& Three linear regression models and three feed-forward neural networks are implemented with an increasing number of input features. Then, the probabilistic forecasts are computed by generating weather scenarios. Residual simulation based on normality assumption is used to improve the probabilistic forecasts.
		The improvement provided by residual simulation is diminishing with the refinement of the underlying model. A very comprehensive underlying model will not benefit from residual simulation based on the normality assumption.
		\\ \hline
		\citet{hyndman2015monash} & $[3,4]$
		&  A methodology to forecast the probability distribution of annual, seasonal and weekly peak electricity demand and energy consumption for various regions of Australia is proposed. The final model is the combination of a short and long-term models. The forecast distributions are derived from the model using a mixture of temperature and residual simulations, and future assumed demographic and economic scenarios.
		Unfortunately, there is no quantitative metric used in this study. However, the twenty-six actual weekly maximum demand values fall within the region predicted from the ex-ante forecast distribution. 
	\end{tabular}
\end{table*}

\newpage


\begin{table*}[!htb]
	\renewcommand\thetable{8}
	\renewcommand{\arraystretch}{1.2}
	\caption{Long-Term Load forecasting summary. \label{tab:LTLF_key_information}} 
	\begin{tabular}{m{1.5cm}m{0.6cm}||m{13.5cm}}
		Author & Class & Forecasting tools and results \\ \hline \hline
		\citet{Hong2008} & $[4,2]$
		&  A forecasting model is composed of a bottom-up and top-down modules. The first one aggregates load for each small area in each level to produce the S-curve parameters. Then, the second one takes these parameters as inputs to allocates the utility's system forecast from the top to the bottom level. A human expert is integrated into the problem solving loop to improve the results from this automated approach.
		The proposed method has been applied to several utilities and has received satisfactory results. 
		\\ \hline 
		\citet{HongXie2014} & $[4,3]$
		& An approach based on linear regression models that takes advantage of hourly information computes long-term probabilistic load forecasts. The short-term load models of \citet{hong2010short} are extended to long-term forecasting by adding the Gross State Product macroeconomic indicator. The probabilistic forecasts are computed by using weather and macroeconomic scenarios. Weather normalization is applied to estimate the normalized load profile. A sliding window methodology is used.
		The customized short-term models performed a MAPE of 4.7 \% for one year ahead forecasting.
		\\ \hline
		\citet{HongShahidehpour2015} & $[4,4]$
		& A review of load forecasting topics and three case studies based on linear regression for long-term load forecasting are exposed. The main effects of the linear regression models are the temperature modeled by a third order polynomial, the calendar variables, and a macro economic indicator. The cross effects are several cross variables of temperature and calendar variables.
		These case studies are exposed to assist Planning Coordinators to provide a real-world demonstration.
	\end{tabular}
\end{table*}

\section{Load forecasting tools}\label{sec:lf_tools}

This paper adopts the distinction made by \citet{Hong2016} between the forecasting techniques and methodologies. A forecasting technique is a group of models that fall in the same family, such as linear regression and artificial neural networks, etc. A forecasting methodology is a framework that can be implemented with multiple forecasting techniques.
Subsections \ref{FT}, \ref{FM}, \ref{annex:DCT}, and \ref{annex:EM} provide some information and references about the forecasting techniques, methodologies, data cleansing techniques and error metrics implemented in the articles reviewed. 

\subsection{Forecasting techniques}\label{FT}

\subsubsection{Basic notions of supervised learning}

\citet{friedman2001elements} provide useful details and explanations about supervised learning algorithms, model assessment and selection. A load forecasting problem is a supervised learning regression problem. In supervised learning, the goal is to predict the value of an outcome measure based on a number of input measures. Supervised learning is opposed to unsupervised learning, where there is no outcome measure, and the goal is to describe the associations and patterns among a set of input measures. A symbolic output leads to a classification problem whereas a numerical output leads to a regression problem. In the statistical literature, the inputs are often called the predictors and more classically the independent variables. In the pattern recognition literature, the term features is preferred. Both terms, predictors or features are used in this paper.

There are three main criteria to compare and select learning algorithms. The accuracy, measured by the generalization error. It is estimated by efficient sample re-use such as cross-validation or bootstrapping. The efficiency which is related to the computation time and scalability for learning and testing. The interpretability related to the comprehension brought by the model about the input-output relationship. Unfortunately, there is usually a trade-off between these criteria. 

\subsubsection{Forecasting techniques implemented}

\begin{figure}[!htb]
	\begin{subfigure}[tb]{0.5\textwidth}
		\includegraphics[width=8cm,height=6cm]{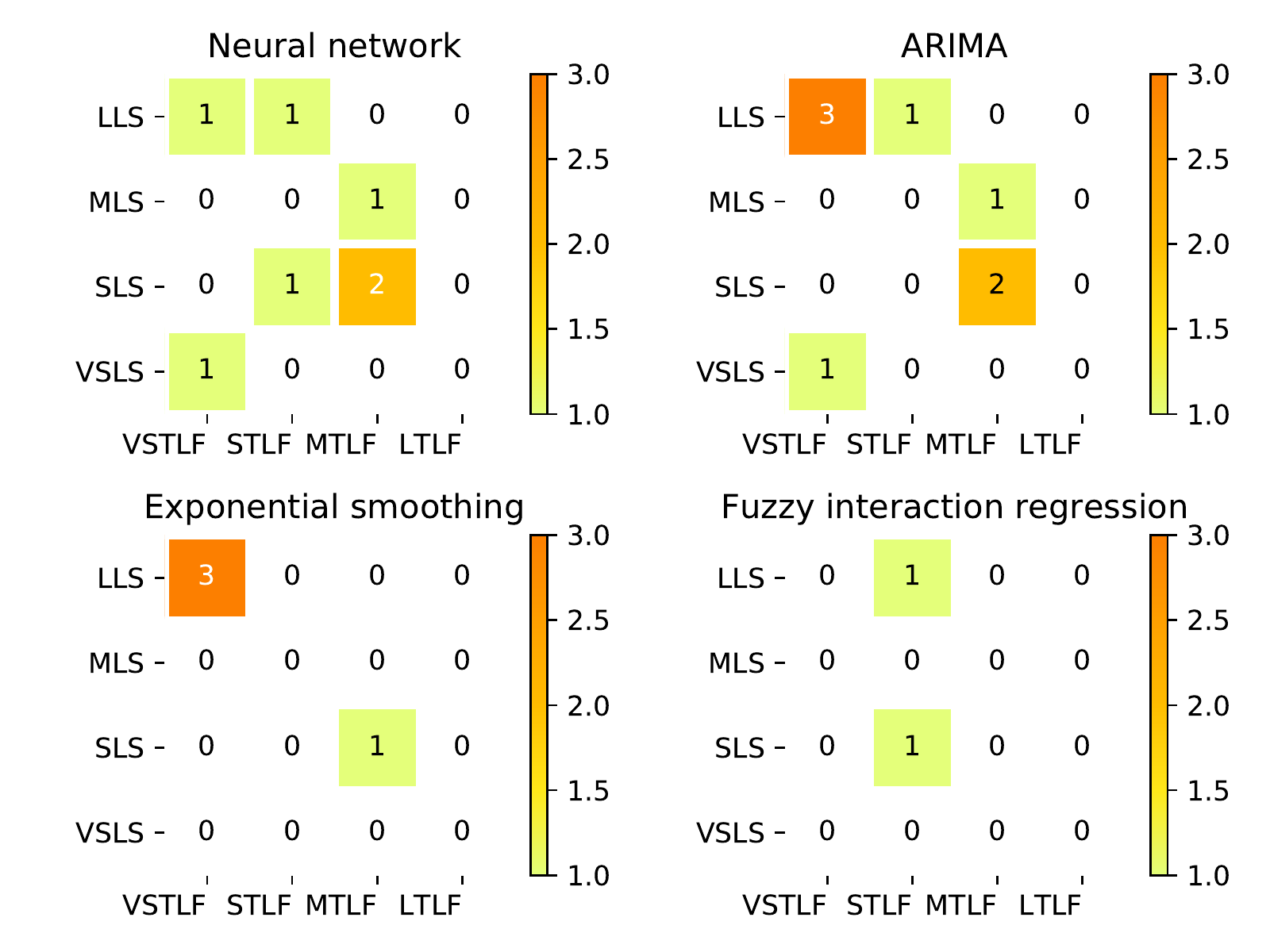}
	\end{subfigure}
	\begin{subfigure}[tb]{0.5\textwidth}
		\includegraphics[width=8cm,height=6cm]{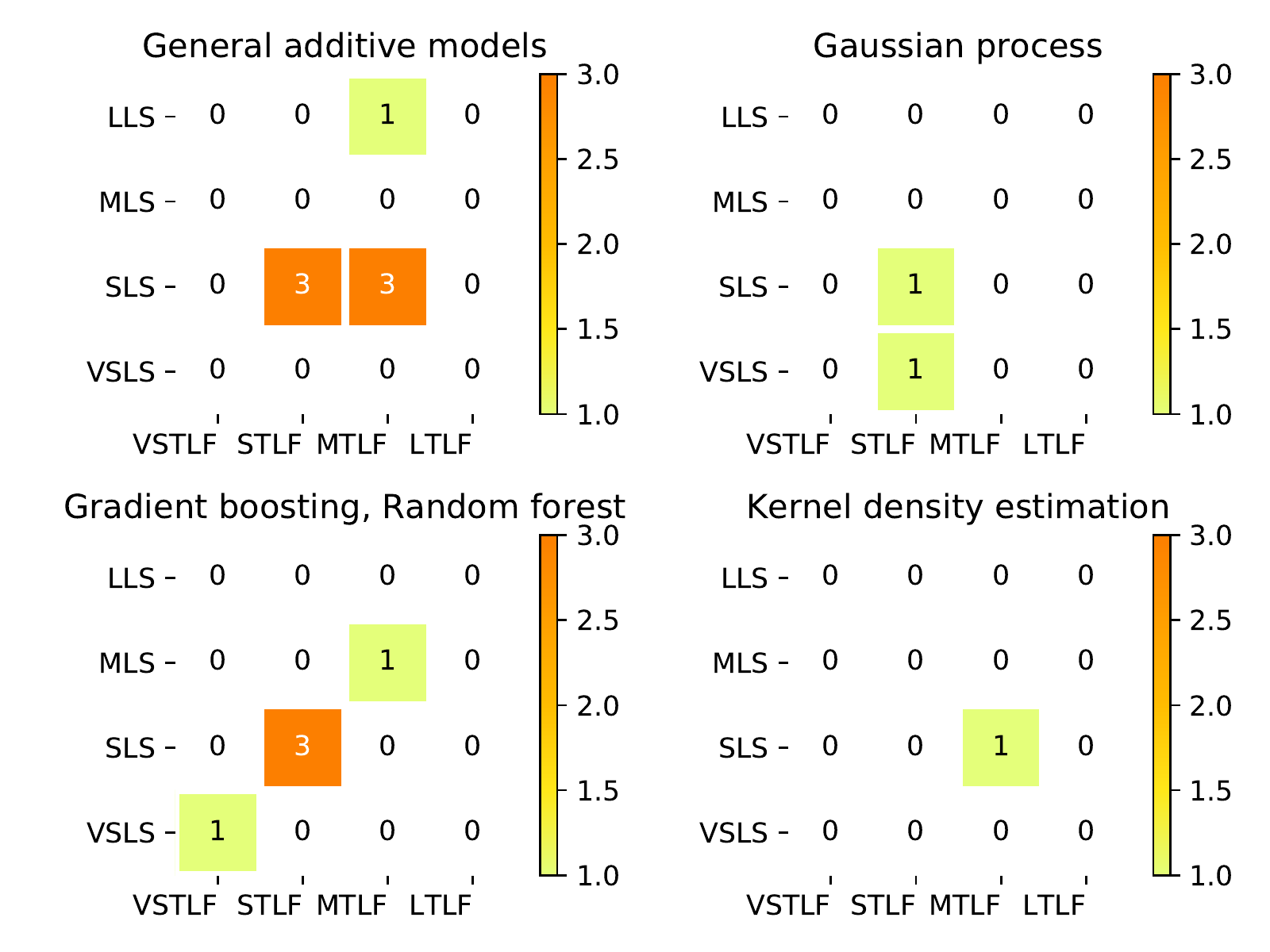}
	\end{subfigure}
	\begin{subfigure}[tb]{0.5\textwidth}
		\includegraphics[width=8cm,height=6cm]{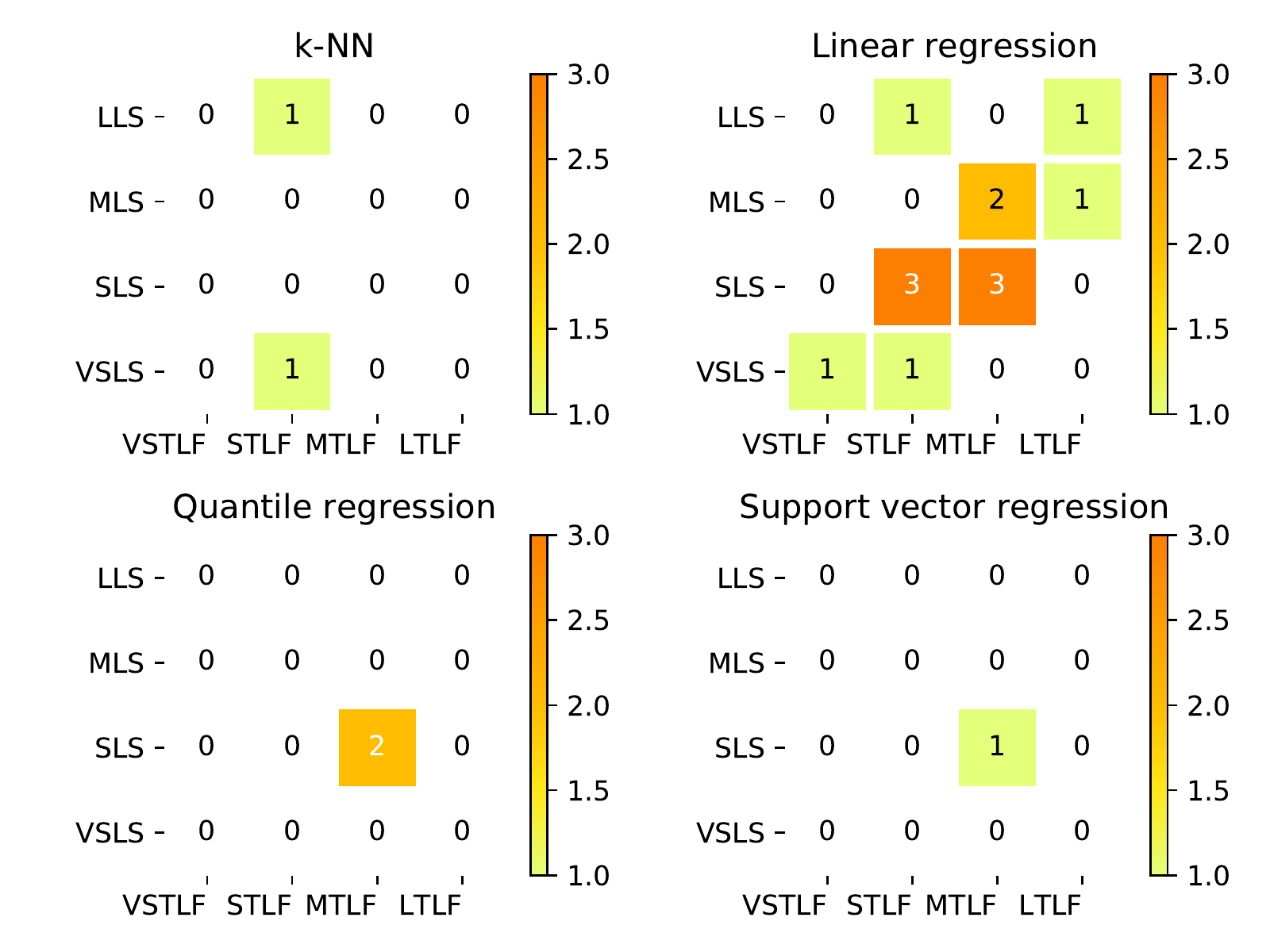}
	\end{subfigure}
	\caption{Load forecasting techniques by forecasting problem.}
	\label{fig:clf_FT_heatmap}
\end{figure}

This subsection provides the basic principles and some references of the forecasting techniques implemented in the articles reviewed. Figure \ref{fig:clf_FT_heatmap} shows for each forecasting technique in the articles reviewed a heat map with the number of implementations by forecasting problem. Without surprising, linear regression, neural networks, gradient boosting and ARIMA are the techniques the most implemented. However, for specific forecasting problems some techniques are more used than other such as exponential smoothing for the class $[1, 4]$.

\paragraph{\textbf{Artificial neural network}}~\\
Artificial neural network basic principles are discussed in \citet{friedman2001elements} and \citet{Weron2006} did a useful review of the various neural network techniques. The artificial neural network is a supervised learning method initially inspired by the behavior of the human brain and consists of the interconnection of several small units. The motivations for neural networks dates back to \citet{mcculloch1943logical} and now the term neural network has evolved to encompass a large class of models and learning methods. 

The central idea of artificial neural network is to extract linear combinations of the inputs as derived features, and then model the target as a nonlinear function of these features. The neural network has unknown parameters, called weights. Their values are calculated by fitting the model to the data with a nonlinear optimization by the back propagation algorithm. Artificial neural networks are universal function approximators. With a sufficient number of neurons and layers they can model any function of the inputs. Artificial neural network are very accurate if the method is well used. However, the learning phase may be very slow, the scalability is not optimal when dealing with large-scale databases. However, the feature selection is an effective tool to improve scalability. Finally, artificial neural networks are black box models and sometimes difficult to interpret. 

\paragraph{\textbf{AR, ARMA, ARIMA and ARX-type models}}~\\
\citet{Weron2006} produced a useful coverage of AR, ARMA, ARIMA and ARX-type models. The standard time series model that takes into account the random nature and time correlations of the phenomenon under study is the Autoregressive Moving Average model (ARMA). It assumes the stationary of the time series under study. As the electricity demand is non-stationary the ARMA models are not always suitable and a transformation of the series to the stationary form is done by differencing. Autoregressive Integrated Moving Average (ARIMA) or Box-Jenkins model, introduced by \citet{box1976time}, is a general model that contained both AR and MA parts and explicitly included differencing in the formulation that are often used for load forecasting. ARX, ARMAX, ARIMAX and SARIMAX are generalized counterparts of AR, ARMA, ARIMA and SARIMA taking into account exogenous variables such as weather or economic.

\paragraph{\textbf{Ensemble methods}}~\\
\citet{friedman2001elements} discuss ensemble methods. The idea is to build a prediction model by combining the strengths of a collection of simpler base models to change the bias variance trade-off. Ensemble methods can be classified into two categories: the averaging and the boosting techniques. The averaging techniques build several estimators independently and then average out their predictions, reducing therefore the variance from a single estimator. Bagging and random forest techniques are part of this family. The boosting techniques involve the combination of several weak models built sequentially. AdaBoost and gradient tree boosting methods fall into this category. The averaging techniques decrease mainly variance whereas boosting technique decreases mainly bias.

\paragraph{\textbf{Exponential smoothing}}~\\ 
\citet{gardner2006exponential} brings the state of the art in exponential smoothing up to date since \citet{gardner1985exponential}. The equations for the standard methods of exponential smoothing are given for each type of trend. Exponential smoothing assigns weights to past observations that decrease exponentially over time. \citet{Hyndman2008} classified fifteen exponential smoothing models versions based on the trend and seasonal components: additive, multiplicative, etc.

\paragraph{\textbf{Fuzzy interaction regression}}~\\ 
\citet{asai1982linear} did the earliest formulation of fuzzy regression analysis. \citet{HongWang2014} define the fundamental difference between the linear regression and fuzzy interaction regression assumptions as the deviations between the observed values and the estimated values. In linear regression, these values are supposed to be errors in measurement or observations that occur whereas they are assumed to depend on the indefiniteness of the system structure in fuzzy interaction regression. \citet{HongWang2014}, \citet{asai1982linear} and \citet{HideoTanakaIsaoHayashi1989} provide useful details and explanations about fuzzy interaction regression theoretical background.

\paragraph{\textbf{Generalized additive models}}~\\ 
Generalized additive models allow the use of nonlinear and nonparametric terms within the framework of additive models and are used to estimate the relationship between the load and explanatory variables such as temperature and calendar variables. According to \citet{friedman2001elements}, the most comprehensive source for generalized additive models is \citet{Hastie1990}. \citet{efron1991statistical} provide an exposition of modern developments in statistics for a nonmathematical audience. Generalized additive models are a useful extension of linear models, making them more flexible while still retaining much of their interpretability.

\paragraph{\textbf{Gradient boosting}}~\\ 
Boosting techniques are part of ensemble methods and discussed by \citet{friedman2001elements}, \citet{Buhlmann2007} and \citet{Schapire2013}.
The motivation for boosting is a procedure that combines the outputs of many weak learners to produce a powerful committee. \citet{schapire1990strength} showed that a weak learner could always improve its performance by training two additional classifiers on filtered versions of the input data stream. Thus, the purpose of boosting is to sequentially apply the weak classification algorithm to repeatedly modified versions of the data, thereby producing a sequence of weak classifiers. Then, the predictions from all of them are combined through a weighted majority vote to produce the final prediction. Boosting regression trees such as Multiple Additive Regression Trees (MART) improves their accuracy often dramatically. However, boosting is more sensitive to noise than averaging techniques (bagging, random forest) and reduces the bias but increases the variance.

\paragraph{\textbf{Gaussian process and Log-normal process}}~\\ 
\citet{Rasmussen2004} defines a Gaussian process as "a collection of random variables, any finite number of which have joint Gaussian distributions". It is fully specified by its mean and covariance functions. This is a natural generalization of the Gaussian distribution whose mean and covariance is a vector and matrix, respectively. The Gaussian distribution is over vectors, whereas the Gaussian process is over functions. The covariance functions, or kernels, encode the relationship between the inputs. The forecasting accuracy is strongly dependent on the kernels selection. \citet{Rasmussen2004} provide a detailed presentation of Gaussian process and their related kernels.

Log-normal process is derived from conventional Gaussian process by performing the logarithm of the normalized load data. Based on the Ausgrid residential dataset (\hyperref[D15]{D15}), \citet{Shepero2018} noticed that the probability distribution of the load is not normally distributed, but is positively skewed and seems to follow a log-normal distribution. \citet{Munkhammar2014} did a similar study based on a residential load dataset by using the Weibull and the log-normal distributions. Thus, depending on the load data distribution, modeling the residential load by a log-normal distribution might provide better results than applying directly Gaussian process.

\paragraph{\textbf{Kernel density estimation}}~\\ 
Kernel density estimation is a nonparametric way to estimate the probability density function of a random variable. A simple kernel density estimate produces an estimate of the probability distribution function of the load using past hourly observations. The kernel function may be Gaussian or of another kind. Kernel density estimation is described in \citet{friedman2001elements} and a good overview of density estimation is given by \citet{silverman2018density} and \citet{scott2015multivariate}.

\paragraph{\textbf{$k$-Nearest Neighbor ($k$-NN)}}~\\
The technique was first introduce by \citet{EvelynFix1951} and later formalized by \citet{Cover1967} for classification tasks. $k$-nearest neighbor regression consists in identifying the $k$ most similar past sequences to the one being predicted, and combines their values to predict the next value of the target sequence. 

\citet{Qahtani2013} review $k$-nearest neighbor for load forecasting with the basic principles. The $k$-nearest neighbor algorithm is specified by four metaparameters: the number $k$ of neighbors used to generate the forecast, the distance function, the operator to combine the neighbors to estimate the forecast and the specification of the univariate or multivariate feature vector including the length of the embedding dimension $m$.
$k$ and $m$ are determined empirically by grid-search over different values depending on the dataset. The most popular distance function is the Euclidean one but other functions might be employed and tested. \citet{Qahtani2013} provide references on this topic. The combination function can employ various metrics such as an equally scheme where neighbors receive equal weight by computing the arithmetic mean or a weighting scheme based on the relative distance of the neighbors. Alternative schemes can be employed and tested such as median, winsorized mean, etc.

$k$-nearest neighbor is a simple algorithm and can be adapted to any data type by changing the distance measure. However, choosing a good distance measure is a hard problem, the algorithm is very sensitive to the presence of noisy variables and is slow for testing.

\paragraph{\textbf{Linear regression}}~\\ 
\citet{friedman2001elements} and \citet{weisberg2005applied} discuss the technique principles. Linear regression analysis is a statistical process for estimating the relationships among variables and one of the most widely used statistical techniques. They are simple and often provide an adequate and interpretable description of how the inputs affect the outputs. The load or some transformation of the load is usually treated as the dependent variable, while weather and calendar variables are treated as independent variables. Linear regression model assumes a linear function of the inputs. They can be applied to transformations of the inputs such as polynomial function of temperature and it considerably expands their scope.

The most popular method to estimate the linear regression parameters is least squares in which the coefficients minimize the residual sum of squares. However, this estimate method often has low bias but large variance that have a direct impact on accuracy. One way to improve it is to use shrinkage methods such as Ridge Regression or Least Absolute Shrinkage and Selection Operator (LASSO), introduced respectively by \citet{hoerl1970ridge} and \citet{tibshirani1996regression}. They shrink the regression coefficients by imposing a penalty on their size.

Linear regression is simple and transparent, there exist fast and scalable variants and provide interpretable models through variable weights (magnitude and sign). However, it is often not as accurate as other (nonlinear) methods.

\paragraph{\textbf{Quantile regression}}~\\ 
Quantile regression was introduced by \citet{Koenker1978} and is a generalization of the standard regression, where each quantile is found through the minimization of a linear model fitted to historical observations according to a loss function. \citet{Gaillard2016} provide a description of this technique.

\paragraph{\textbf{Regression tree and Random forest}}~\\ 
\citet{friedman2001elements} provide useful details and explanations about regression tree and random forest. \citet{breiman1984classification} introduced the Classification and Regression Trees (CART) methodology and \citet{quinlan1986induction} gives an induction of regression tree. A decision tree is a tree where each interior node tests a feature, each branch corresponds to a feature value and each leaf node is labeled with a class. Tree for regression are exactly the same model than decision tree but with a number in each leaf instead of a class. A regression tree is a piecewise constant function of the input features. Overfitting is avoided by pre-pruning (stop growing the tree earlier, before it reaches the point where it minimizes the training error, usually the squared error, on the learning sample), post-pruning (allow to overfit and the tree that minimizes the squared error on the learning set is selected) or by using ensemble method: random forest, boosting. Regression tree is a very fast and scalable technique (able to handle a very large number of inputs and objects), provides directly interpretable models, and gives an idea of the relevance of features. However, it has a high variance and is often not as accurate as other techniques. \\

\citet{Breiman2001} defines random forest as a combination of tree predictors such that each tree depends on the values of a random vector sampled independently and with the same distribution for all trees in the forest. The random forest method is a substantial modification of bagging that builds a large collection of de-correlated trees, and averages them to reduce the variance. Random forest combines bagging and random feature subset selection. It builds the tree from a bootstrap sample and instead of choosing the best split among all features, it selects the best split among a random subset of $k$ features. There is a bias variance trade-off with $k$. The smaller $k$ is, the greater the reduction of variance is but also the higher the increase of bias is. Random forest has the advantage to decrease computation time with respect to bagging since only a subset of all features needs to be considered when splitting a node.

\paragraph{\textbf{Support vector regression}}~\\ 
The theory behind support vector machines is due to \citet{vapnik2013nature}. The support vector machine produces nonlinear boundaries by constructing a linear boundary in a large, transformed version of the feature space. This technique is based on two smart ideas: large margin classifier and kernelized input space. The margin is the width that the boundary could be increased by before hitting a data point. Linear support vector machine is the linear classifier with the maximum margin. If data is not linearly separable the solution consists in mapping the data into a new feature space where the boundary is linear. Then, to find the maximum margin model in this new space. In fact, there is no need to compute explicitly the mapping. Only a similarity measure between objects (like for the $k$-nearest neighbor) is required. This similarity measure is called a kernel. This procedure is sometimes called the kernel trick.

Support vector regression uses the same principles as the support vector machine for classification. Support vector machines are state-of-the-art accuracy on many problems and can handle any data types by changing the kernel. However, tuning the method parameter is very crucial to get good results and somewhat tricky, it is a black-box models and not easy to interpret. \citet{Cortes1995}, \citet{Drucker1997} and \citet{friedman2001elements} provide details and explanations about support vector machines for classification and regression.

\paragraph{\textbf{Unobserved components model}}~\\ 
Unobserved component model, introduced in \citet{harvey1990forecasting}, decomposes a time series into trend, seasonal, cyclical, and idiosyncratic components and allows for exogenous variables. Unobserved component model is an alternative to ARIMA models and provides a flexible and formal approach to smoothing and decomposition problems.

\subsection{Forecasting methodologies implemented}\label{FM}

\begin{figure}[!htb]
	\begin{subfigure}[b]{0.5\textwidth}
		\includegraphics[width=8cm,height=6cm]{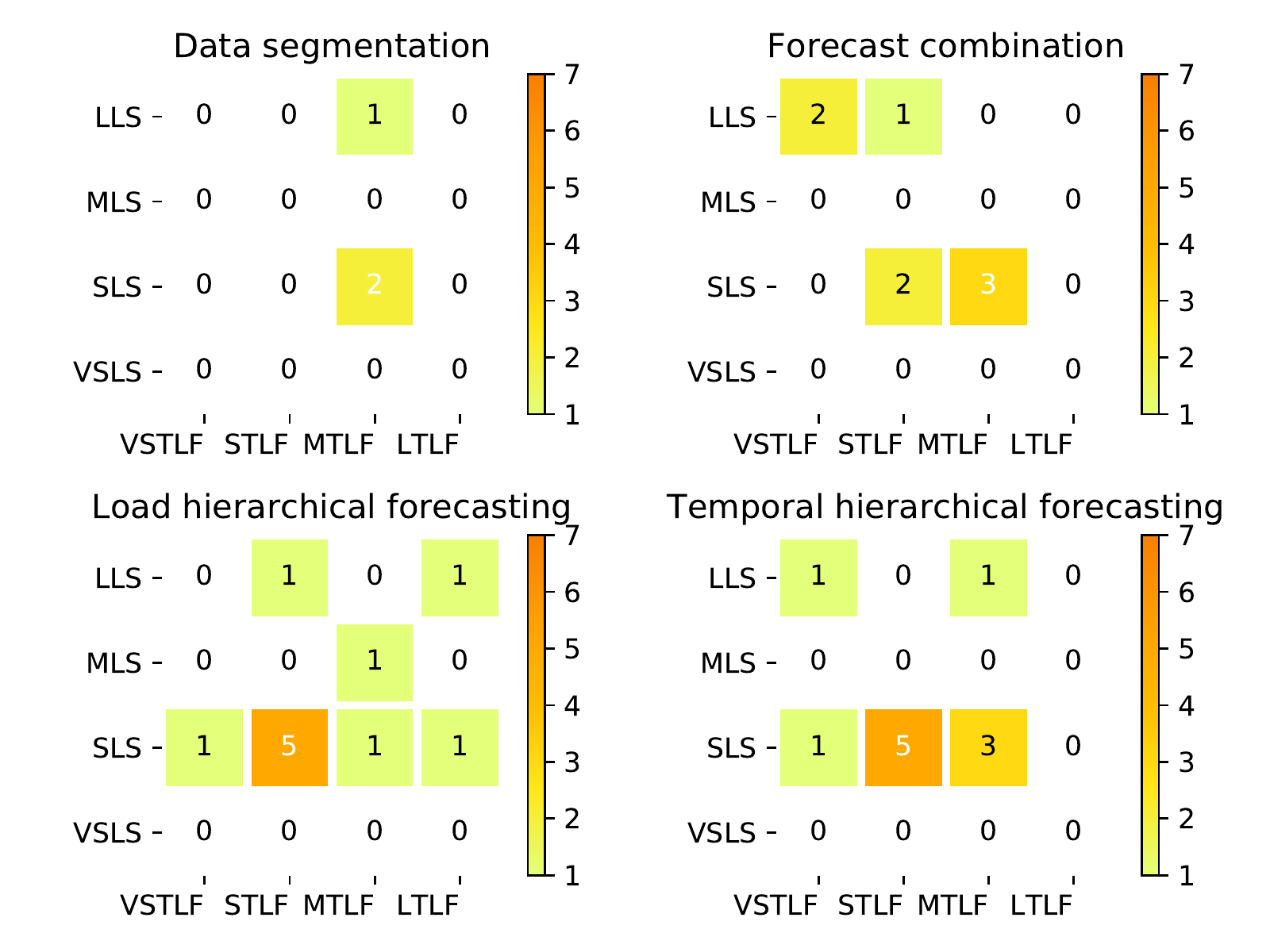}
	\end{subfigure}
	\begin{subfigure}[b]{0.5\textwidth}
		\includegraphics[width=8cm,height=6cm]{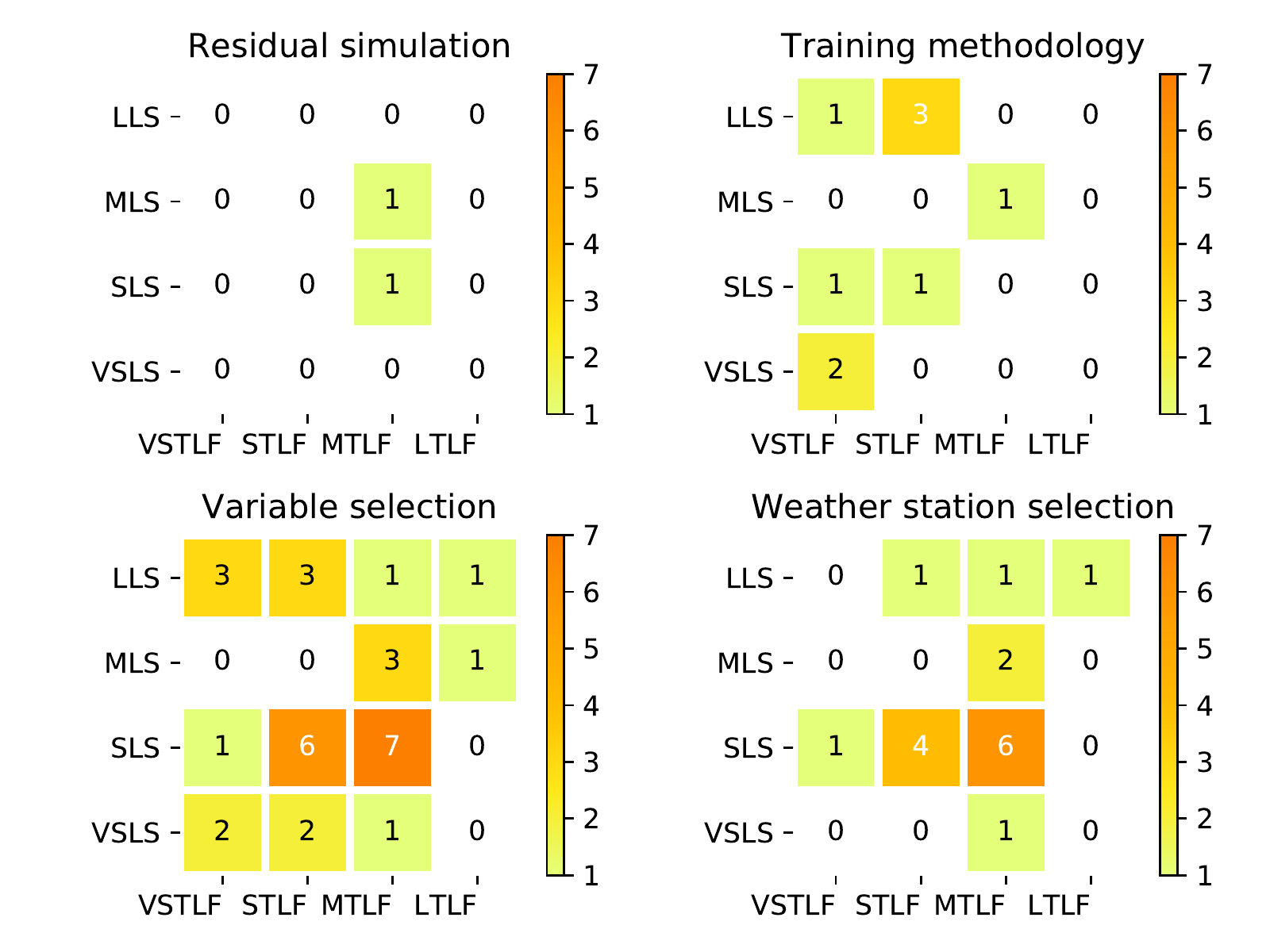}
	\end{subfigure}
	\caption{Load forecasting methodologies by forecasting problem.}
	\label{fig:clf_FM_heatmap}
\end{figure}

This subsection provide the basic principles and references of the forecasting methodologies implemented in the articles reviewed. Figure \ref{fig:clf_FM_heatmap} shows for each forecasting methodology a heat map with the number of implementations by forecasting problem.

\paragraph{\textbf{Data Segmentation (DS)}}~\\ 
Data segmentation consists in slicing the dataset into several parts and training one or several models on each one or some of them. Then, these models produce forecasts on the corresponding segments where they have been trained. For instance, \citet{Chen2001} used only the winter data segment to forecast the January load with support vector regression.

\paragraph{\textbf{Forecast Combination (FC)}}~\\  
Many authors have suggested the superiority of forecast combinations over the use of individual forecast. \citet{hibon2005combine} developed a simple model-selection criterion. The accuracy of the selected combination was significantly better and less variable than the selected individual forecasts. 

Forecast combination is similar to the bagging technique. Bagging (bootstrap aggregating) uses bootstrap sampling to generate several learning samples, train the model on each one of them and compute the average. Variance is reduced but bias increases a bit because the effective size of a bootstrap sample is about 30 \% smaller than the original learning set. However, forecast combination differs from bagging as it consists in combining predictions of different models. There exist several ways of combining forecasts. 

Forecast Simple Averaging Combination (FSAC) is the most trivial and consists simply in averaging the forecasts. Forecast Weighted Combination (FWC) is more refined and consists in using a weighted average. The weights are calculated with different methodologies: manually, by solving an optimization problem, etc. \citet{Nowotarski2014} and \citet{Nowotarski2016} developed alternative schemes for combining forecasts: simple, trimmed averaging, winsorized averaging, ordinary least squares, least absolute deviation, positive weights averaging, constrained least squares, inverse root mean squared error, Bayesian model averaging, exponentially weighted average, and fixed share machine learning techniques, polynomial weighted average forecasting technique with multiple learning rates, best individual ex-ante model selection in the validation period and best individual ex-ante model selection in the calibration window. \citet{Ranjan2010} developed a beta-transformed linear opinion pool for the aggregation of probability forecasts from distinct, calibrated or uncalibrated sources. 

\paragraph{\textbf{Load Hierarchical forecasting (LHF)}}~\\ 
Load hierarchical forecasting consists in adopting a spatial approach to improve forecasts at the aggregated and local levels. A load system can be divided into several zones with different load patterns. Residential zones are more sensitive to temperature effect and industrial zones are more sensitive to economic parameters or workload. These zones are modeled separately with different forecasting techniques and methodologies. Then, the system load is forecasted by aggregating all the forecasts at the zonal level. 

The Global Energy Forecasting Competition 2012 load track \citet{hong2014global} is an example of load hierarchical forecasting problem with a US utility composed of twenty zones. The forecasting task was to predict the load value at both the zonal (twenty series) and system (sum of the twenty zonal level series) levels.

\paragraph{\textbf{Residual Simulation (RS)}}~\\
Residual simulation is a way to produce probabilistic load forecasting by post-processing the point forecasts \citet{Hong2016}. Applying the probability density function of residuals to the point forecasts generates a density forecast. The normality assumption is often used to model the forecasting errors. \citet{Xie2017} investigated the consequences of this assumption and showed that it helps to improve the probabilistic forecasts from deficient underlying models but the improvement diminishes as the underlying model is improved. It confirms the importance of sharpening the underlying model before post-processing the residuals. However, if it is not feasible to build comprehensive underlying models, modeling residuals with normal distributions is a plausible method. 

\paragraph{\textbf{Similar day (SD)}}~\\
The "similar day" approach considers a "similar" day in the historic data to the one being forecast. Due to its simplicity, it is regularly implemented in industrial applications. The similarity is usually based on calendar and weather patterns. The forecast can be a linear combination or a regression procedure that include several similar days. \citet{Taylor2008} implemented a development of this idea in a weather-based forecasting approach which is described by \citet{taylor2003using}.

\paragraph{\textbf{Temporal Hierarchical forecasting (THF)}}~\\ 
\citet{Athanasopoulos2017} introduced the concept of temporal hierarchies for time series forecasting, using aggregation of non-overlapping observations. By combining optimally the forecasts from all levels of aggregation, this methodology leads to reconciled forecasts supporting better decisions across planning horizons, increased forecast accuracy and mitigating modeling risks. \citet{Gaillard2016} concatenated a short-term (from one hour to forty-eight hours) and a medium-term (from forty-nine hours to one month) probabilistic forecasting models. \citet{Nedellec2014} developed a temporal multi-scale approach by modeling the load with three components: a long, medium, and short-term parts.

Temporal hierarchical forecasting consists also in developing forecasting models for specific time periods. Models for specific hours of the day: one per hour, one per night hours, one for morning hours, one for afternoon, and one for evening hours, etc. Models for specific days: one per week days, one per weekend days, etc. Models for specific time of the year: one per month, one per season, etc. 

\paragraph{\textbf{Training Methodologies (TM)}}~\\ 
The way to train a model has a deep impact on the results. Several methodologies exist depending on the past data available and the newly data acquired during the forecast process. Static training methodology consists in training once for all the model with all, or a segment, of data available. Then, it produces forecasts without retraining if newly data are acquired. Accumulative training methodology consists in retraining periodically the model by using newly data acquired and the past data. The retraining period depends on the forecasting horizon. Usually, the smaller the forecasting horizon is, the higher the retraining frequency is. Sliding window methodology consists in using a fixed training data window that is shifted periodically using the newly accumulated data. The shifting period depends on the forecasting horizon. Usually, the smaller the forecasting horizon is, the higher the shifting period frequency is.

\paragraph{\textbf{Variable selection (VS)}}~\\ 
Variable selection assesses the features to use and their functional forms. The goal is to find a small, or the smallest, subset of features that maximizes accuracy. This process enables to avoid overfitting and improves the model performance and the interpretability. It also provides faster and more cost-effective models and reduces the overall computation time. \citet{guyon2003introduction,guyon2006introduction,saeys2007review}, and \citet{friedman2001elements} are useful references about feature selection. Three main approaches exist for variable selection. 

The first one is the filter technique and consists in selecting the relevant features with methods independent of the supervised learning algorithm implemented. The univariate statistical tests (t-test, chi-square, etc.) are fast and scalable but ignore the feature dependencies. The multivariate approaches (decision trees, etc.) take into account the feature dependencies but are slower than univariate. The cross-validation is a powerful tool but can be computationally expensive.

The second approach is the embedded technique. The search for an optimal subset of features is sometimes already built into the learning algorithm. Decision tree node splitting is a feature selection technique, tree ensemble measures variable importance, the absolute weights in a linear support vector machine model, and a linear model with Least Absolute Shrinkage and Selection Operator provide a feature selection. The embedded technique is usually computationally efficient, well integrated within the learning algorithm and multivariate. However, it is specific to a given learning algorithm.

The last approach is the wrapper technique. It tries to find a subset of features that maximizes the quality of the model induced by the learning algorithm. The quality of the model is estimated by cross-validation. All subsets cannot be evaluated and heuristics are necessary as the number of subsets of $p$ features is $2^p$. Several approaches exist. Among them, the forward (or backward) recursive feature elimination consists in adding (removing) the variable that most decreases (less increases) the error over several iterations. The wrapper technique is custom-tailored to the learning algorithm, able to find interactions and to remove redundant variables. However, it is prone to overfitting. Indeed, it is often easy to find a small subset of noisy features that contribute to decreasing the score. In addition, this technique is computationally expensive as a model is built for each subset of variables.

Almost all of the articles reviewed adopt a variable selection approach. \citet{hong2010short} developed a methodology to select the relevant features for linear regression, fuzzy interaction regression and artificial neural network techniques. The increasing number of relevant features improved the forecast accuracy. \citet{HongXie2014} extended the method to long-term load forecasting by adding a macroeconomic indicator. As in \citet{hong2010short}, \citet{Charlton2014} did a series of refinements to the linear regression benchmark proposed during the Global Energy Forecasting Competition 2012: day-of-season terms, a special treatment of public holidays, and changing the number of seasons. The competition scores were used to demonstrate that each successive refinement step increases the model accuracy.

\paragraph{\textbf{Weather station selection (WSS)}}~\\
The weather is a major factor driving the electricity demand, price, wind or solar power generation. \citet{Pang2012} details two methodologies to capture the weather impact. The Virtual Weather Station methodology (VWS) consists in a combination (simple average, weighted average, etc.) of the parameters (temperature, wind speed, solar irradiation, etc.) of several weather stations. At the end of the process, the temperature (or solar irradiation, wind speed, etc.) combination can be seen as a parameter from a virtual weather station and is used as a feature for the forecasting models. The Multiple Weather Station (MWS) consists in selecting the weather stations, whose parameters, improve the most the model accuracy. Each one of them is used as feature to feed a forecasting model. The final forecast is the combination (simple average, weighted average, etc.) of the forecasts generated by each model. Based on this principle, \citet{hong2015weather} developed a framework to determine how many and which weather station selecting for a territory of interest.

\section{Use case example and guidelines}\label{sec:guidelines}

Sections \ref{sec:lf_classification} and \ref{sec:lf_tools} present general guidelines for selecting which forecasting technique or methodology to use in which case. However, the classification process in two dimensions does not take into account the shape or frequency of the load, the quality or length of the data, at which frequency the data are available, etc. Each forecasting problem is thus specific and requires a dedicated customized forecasting model. This section proposes some guidelines, and emphasizes some difficulties that may be encountered to build a first model. It is exemplified on the methodology used in \citet{dumas2020coordination}. The text in italic refers to this case.

\subsection{Guidelines}
\begin{enumerate}
	\item Understand and formulate the forecasting problem. What is the application? What are the goals? What is the system considered? \textit{Forecasting the consumption and PV production of a microgrid to provide reliable inputs to a model predictive control to manage in real time the microgrid.}
	\item What are the forecasting horizon, the temporal and spatial resolutions required? \textit{The decision making module solves each quarter an optimization problem for the twenty-four hours ahead with a fifteen minutes resolution. The spatial resolution is in kW. Thus, the forecasting model should at least each quarter provides a forecast for the twenty-four hours ahead with a fifteen minutes resolution and a spatial resolution in kW.}
	\item Is probabilistic forecasting required? \textit{No, as the decision making module is deterministic in this case.}
	\item Which error metrics should be selected to assess the forecasts? \textit{The normalized root mean squared and mean absolute errors are selected as we consider only deterministic forecasts.}
	\item What are the data available? At which frequency is the database updated? What is the quality of the data? \textit{The PV and consumption are monitored on site each five seconds. Weather forecasts are available every six hours for the next ten days with a fifteen minutes resolution. Missing data occur rarely.}
	\item Is data cleansing needed? \textit{Missing data are replaced by zero. However, it occurs rarely.} 
	\item What segment of data is considered for the study?\textit{ Two months of monitored consumption, production, and weather forecasts.}
	\item Is a benchmark model defined?\textit{ A linear regression model could have been implemented as benchmark. However, this study focuses on neural networks and gradient boosting regression.}
	\item What are the forecasting techniques selected?\textit{ A gradient boosting regression model, a feed-forward and a recurrent neural networks.}
	\item Is load hierarchical forecasting relevant?\textit{ No, the microgrid is only composed of one load and the PV monitored is the total production.}
	\item Is temporal hierarchical forecasting relevant?\textit{ The forecasts are composed each quarter of ninety-six values as the horizon is twenty-four hours and the resolution fifteen minutes. Thus, it may be interesting to decompose the horizon in two or four parts with a dedicated model for each one. Then, the final forecast could be a combination of each term. However, for the sake of simplicity, the temporal hierarchical forecasting methodology is not implemented.}
	\item Is forecast combination relevant? \textit{It may be interesting to combine neural networks and gradient boosting models. However, for the sake of simplicity, each model is considered independently.}
	\item What training methodology to use? \textit{A sliding window methodology. The learning set has a size of one week and the model is retrained each day. }
	\item Is weather station selection relevant? \textit{No, the models take as inputs the weather forecasts computed at a location close to the microgrid (less than two km).}
	\item What are the relevant variables of this forecasting problem? \textit{Several sets of variables have been tested. The final set is composed of lags of load and temperature weather forecast for the prediction of the consumption. }
	\item What are the relevant hyper parameters of the models? \textit{Several sets of hyper parameters have been tested. }
\end{enumerate}

\subsection{Forecasting techniques}
In some cases, one or several of these general forecasting techniques guidelines may be of good use:
\begin{enumerate}
	\item Define a benchmark model with a transparent, simple and familiar forecasting technique. Linear regression is recommended for this purpose, see \citet{hong2010short,hong2015weather,Wang2016,Charlton2014,Xie2015LTLF,Xie2015,HongXie2014}. It is easy to implement, transparent and fast to compute in comparison with more complex techniques such as artificial neural networks, Gaussian process or gradient boosting regression. In addition, customized linear regression models may provide interesting results from very short-term to long-term forecasting. \citet{Charlton2014} won the load track of the Global Energy Forecasting Competition 2012 with a customized linear regression model.
	\item Define a robust methodology to find out the relevant hyper parameters of a given forecasting technique. A blocked form of $k$-fold cross-validation procedure \citet{bergmeir2012use} is recommended when the dataset has a minimum length.
	\item Statistical models such as ARIMA or exponential smoothing may provide satisfactory results for horizons of a few hours, where the weather has not a significant impact \citet{Taylor2008}. When the horizon is extending to a few days it is recommended to assess the importance of exogenous parameters such as the temperature.
	\item Gaussian process provides directly probabilistic forecasts. However, the selection of the relevant kernels is not straightforward and the time computation may be an issue depending on the forecasting problem. In addition, forecasting directly multiple outputs is non-trivial and still a field of active research \cite{wang2015gaussian} and \cite{liu2018remarks}. \citet{VanderMeerShepero2018} used a moving training window to reduce the learning set size and consequently the computational burden of the Gaussian process.
	\item Probabilistic forecasting models that produce constant prediction intervals over time provide not useful information for decision making. Indeed, they do not differentiate periods of high and low uncertainty. 
	\item Residual simulation based on normality assumption provides a way to generate probabilistic forecasts \citet{xie2016,Xie2017}. However, this assumption must be assessed. \citet{Xie2017} demonstrated that the improvement provided by residual simulation based on the normality assumption is diminishing with the improvement of the underlying model.
\end{enumerate}

\subsection{Forecasting methodologies}
In some cases, one or several of these general forecasting methodologies guidelines may be of good use.
\begin{enumerate}
	\item Use a blocked form of cross-validation for time series evaluation \citet{bergmeir2012use,VanderMeerShepero2018} instead of a traditional $k$-fold cross-validation. Indeed, the consumption of a utility is a process that might evolves over time, thus hurting the fundamental assumptions of cross-validation that the data are independent and identically distributed.
	\item Random forest, Gradient boosting regression, and LASSO are techniques that perform feature extraction. \citet{hong2010short} developed a methodology to select the relevant features for linear regression, fuzzy interaction regression and artificial neural network techniques. It can be extended to other techniques.
	\item The sliding window training methodology is a way to speed up the time computation by reducing the learning set size \citet{VanderMeerShepero2018}. However, the accuracy may decrease accordingly. A trade-off should be reached between time computation and accuracy. 
	\item In some cases, temporal hierarchical forecasting does not improve the results \citet{Wang2016,hong2010short}. At the opposite the case studies of \citet{Charlton2014,Nedellec2014,bentaieb2013,Goude2014,Gaillard2016,Haben2016} benefited from this methodology. The higher the horizon is and consequently the number of points to forecast, the more it is interesting to consider this methodology.
	\item The same observation can be done with load hierarchical forecasting. \citet{hong2014global} built a case study (\hyperref[D1]{D1}) to experiment this methodology. Large systems such as distribution networks that are composed of hundreds or thousands of substations such as the case study of \citet{Goude2014} may benefit from this methodology to take into account specific locational effects.
	\item The weather station selection framework of \citet{hong2015weather} is recommended when dealing with several possibilities for one location.
	\item In general, a combination of models improves the results such as in \citet{Taylor2008,Lloyd2014,Lopez2018,Xie2015,xie2016,Haben2016}.
	\item Transforming the data by taking the logarithm before applying Gaussian process may improve the results when considering residential consumption as it is better represented by the log-normal distribution \citet{Munkhammar2014, Shepero2018}.
	\item Probabilistic load forecasting is a specific topic. The Global Energy Forecasting Competition 2014 load track \citet{hong2016probabilistic} provides a case study (\hyperref[D6]{D6}) to experiment techniques and methodologies to addressed this problem. \citet{xie2016} propose an integrated solution composed of pre-processing, forecasting and post-processing to compute probabilistic forecast based on several forecasting techniques.
	\item Temperature is often used as input of load forecasting model. However, a temperature forecast with small accuracy may decrease the model performance \citet{Chen2001}.
\end{enumerate}

\section{Conclusion}

The key contribution of this paper is to propose a classification into two dimensions of the load forecasting studies to help to decide which forecasting tools to use in which case. This process aims to provide a synthetic view of the relevant forecasting techniques and methodologies by forecasting problem. 

The two-dimensional classification methodology is based on the definition of a forecasting problem with the forecasting horizon and resolution, the system size and the load resolution. Each article is classified with key information about the dataset used and the forecasting tools implemented: the forecasting techniques (probabilistic or deterministic) and methodologies, the data cleansing techniques, and the error metrics. 

This paper can be read in two passes. The first one by identifying the forecasting problem of interest to select the corresponding class in Tables \ref{tab:VSTLF} to \ref{tab:LTLF}. They provide a synthetic view of the forecasting tools used. Then, key information about the forecasting tools and the results of these studies are summarized in Tables \ref{tab:VSTLF_key_information} to \ref{tab:LTLF_key_information}. The second pass consists in reading the key principles of the main techniques and methodologies of interest and the reviews of the papers in \ref{annex:Review}.

In the future it would be interesting to review more studies classified in the classes [long-term load forecasting, from very small to large load systems] and [from very short-term to long-term load forecasting, medium load system] in order to gather forecasting tools related to these forecasting problems. 
However, to keep a readable and concise document, an extension of this work could be a kind of database where only synthetic information about the forecasting tools and key results are stored in the format of Tables \ref{tab:VSTLF} to \ref{tab:LTLF}. Indeed, the more studies are classified the more forecasting tools are available and the easier it is to select the relevant forecasting tools according to a specific forecasting problem. Then, the statistics provided by Figures \ref{fig:clf_FT_heatmap} and \ref{fig:clf_FM_heatmap} about which forecasting techniques or methodologies used in which case would be more relevant. Thanks to this database, a rapid search within a given forecasting problem would point out a set of potential forecasting tools and a list of references to investigate further their implementations.


\bibliographystyle{elsarticle-num-names}
\bibliography{manuscript}

\twocolumn

\appendix

\section{Notation}
\label{annex:notation}

\noindent \textbf{\textit{Acronyms}}

\renewcommand{\arraystretch}{1.1}
\begin{supertabular}[tb]{m{1.5cm}l}
	Name & Description \\ \hline
	FP & Forecasting Problem \\
	PF & Probabilistic Forecasting \\
	EM & Error metric \\
	DC & Data cleansing \\
	FM & Forecasting methodology \\
	FT & Forecasting technique \\
\end{supertabular}
\\\\
\textbf{\textit{Classification criteria}} 

\begin{supertabular}[tb]{m{1.5cm}l}
	Name & Description\\ \hline
	$\Delta_t$ & Forecasting resolution \\
	$H_{T}$ & Forecasting horizon \\
	VSTLF & Very short-term load forecasting  \\
	STLF &  Short-term load forecasting  \\
	MTLF & Medium-term load forecasting  \\
	LTLF & Long-term load forecasting  \\
	$\Delta_L$ & Spatial resolution of the forecasts\\ 	
	$H_{L}$ & System size \\
	VSLS & Very small load system  \\
	SLS & Small load system  \\
	MLS & Medium load system  \\
	LLS & Large load system  \\
\end{supertabular}
\\\\
\textbf{\textit{$H_{T}$ units}} 

\begin{supertabular}[tb]{m{1.5cm}l}
	Name & Description \\ \hline
	m, h & minute, hour  \\
	d, w & day, week  \\
	M, y & month, year  \\
\end{supertabular}
\\\\

\newpage
\noindent \textbf{\textit{Forecasting methodologies}} 

\begin{supertabular}[tb]{m{1.5cm}l}
	Name & Description \\ \hline
	DS & Data segmentation \\
	FC & Forecast combination \\
	FSAC & Forecast simple average combination \\
	FWC & Forecast weighted combination \\
	LA & Local averaging \\
	LHF & Load hierarchical forecasting \\
	MWS & Multiple weather stations \\
	RS & Residual simulation \\
	THF & Temporal hierarchical forecasting \\
	TM & Training methodologies \\
	VS & Variable selection  \\
	VWS & Virtual weather station \\
	WSS & Weather station selection \\
\end{supertabular}
\\\\
\textbf{\textit{Error Metric}} 

\begin{supertabular}[tb]{m{1.5cm}l}
	Name & Description \\ \hline
	APE & Absolute percentage error  \\
	MAE  & Mean absolute error \\
	MAPE &  Mean absolute percentage error \\
	NRMSE & Normalized root mean square error \\
	PICP &  Prediction Interval Coverage Probability\\
	PINAW &  Prediction Interval Normalized Average Width\\
	PLF &  Pinball loss function\\
	RMSE & Root mean square error \\
\end{supertabular}

\onecolumn

\section{Review section}\label{annex:Review}

\subsection{Very Short-Term Load Forecasting}\label{annex:Review_VSTLF}

\textbf{\citet{VanderMeerShepero2018}} propose Gaussian process to compute probabilistic forecast of a residential building demand, PV production, and net demand. The data cleansing technique used is manual threshold.
An important issue of the Gaussian process is the computation time to learn the hyper parameters of the covariance functions. The dynamic Gaussian process approach is implemented to reduce the computational burden to produce on-line forecasts by using a moving training window. This approach is compared to static Gaussian process and ARIMA models. Several sets of demand lags and different combinations of covariance functions are investigated for both Gaussian process approaches. The linear combination of the squared exponential and Mat\'ern kernels demonstrated the best results. Deterministic and probabilistic metrics are computed using a blocked form of $k$-fold cross-validation procedure.
The MAPE and NRMSE of both Gaussian process models are within the intervals $[3,3.9]$ \% and $[5,8.2]$ \%. MAPE and NRMSE values of ARIMA are smaller and are within the intervals $[1.2,1.8]$ \% and $[1.8,2.5]$ \%. However, ARIMA produced higher prediction intervals that remain constant over time expressing the same certainty for different periods of the day or month. This could be problematic for decision making. In contrast, the Gaussian process had wide and narrow prediction intervals between periods of high and low uncertainty.
The net demand forecasting is achieved by a direct and indirect strategy. The latter subtracts the individual forecasts to produce the net demand forecast. The static Gaussian process was generally better able to capture the peaks than the dynamic Gaussian process approach, while the latter produces more narrow prediction intervals. Overall, selecting the best strategy between direct or indirect depends mainly on whether one prefers higher informativeness of prediction intervals or higher coverage probability. For both Gaussian process approaches, the NMAE and NMRSE of the indirect strategy are smaller than the direct one. \\

\textbf{\citet{Chae2016}} use feature extraction with random forest and artificial neural network to compute a quarterly day-ahead forecast of a building electricity demand. The data are provided by the building management system and consists in instantaneous power electricity measured in kW with a minute interval that are aggregated each quarter in kWh. As data cleansing, four outlier days are removed due to electric meter failure. 
The random forest feature extraction consists in ranking nine potential predictors divided in three categories: the environment variables, the time indicators, and the operational conditions. The first five highly ranked predictors are selected.
Then, nine machine-learning algorithms are assessed based on correlation coefficient and coefficient variance of root mean square error CV(RMSE), that is a normalized RMSE. The artificial neural network was selected. It is a one hidden layer feed-forward network using a back-propagation algorithm.
Three training methodologies are implemented: static, accumulative, and sliding window. The training data length is also assessed and the results indicated that the larger the training dataset is, the more accurate the model is.
The accumulative and sliding window training methodologies produced the best results in terms of daily averaged CV(RMSE): approximately 8-9 \% and 10 \% for weekdays and weekends. The static training method produced results from 9 \% to 14 \% for weekdays and 11 \% to 27 \% for weekends. All training methods predicted the daily peak demand with an APE of approximately 3 \% and 4.5 \% for weekdays. Concerning weekends, the results depend on the month considered. These results are promising and the authors claim this methodology could be used for a model predictive control to reduce energy costs in buildings. \\

\textbf{\citet{Wang2016}}\label{12Wang2016} use a linear regression model and a case study based on data (\hyperref[D1]{D1}) from the load forecasting track of the Global Energy Forecasting Competition 2012 \citet{hong2014global} to study the recency effect. This investigation is done both at the aggregated and bottom levels and for each hour of the day. Modern computing power is used to develop forecasting models with thousands of variables to customize the recency effect, for each zone and each hour of the day.
The recency effect is modeled by a combination of lagged hourly temperatures and/or moving average temperatures. It is an extension of the method proposed in~\citet{hong2010short}. At the aggregated level, the pair with two daily moving average temperatures and twelve lagged hourly temperatures provide the best MAPE. At the bottom level, the recency effect appears to differ across most of the nineteen zones. 
Two zones are listed separately due to their load temporal behaviors, one experienced a major outage and the other one is an industrial customer. However, no data correction or removal is done.
Two training methodologies are considered. A static methodology where the dataset is sliced into three pieces with the first two years for training, the third for validation, and the last year for testing. A sliding window methodology using a 730-day moving window for daily forecasts on rolling basis. Finally, the weather station selection framework of \citet{hong2015weather} is used to compute virtual weather stations.
The customized linear regression model with the recency effect is compared to the benchmark model (GLMLF-B7) proposed by~\citet{hong2010short} and used in \citet{hong2014global, hong2015weather}. The results indicate that developing a model per hour, by slicing the data into twenty-four pieces, is not necessarily better than one model for all the hours of the day. 
When considering the forecasting horizon of one year, at the aggregated level, the recency effect reduces the MAPE by 18 \% from 5.22 \% to 4.27 \%. At the bottom level, the MAPE values are reduced for seventeen of the eighteen regular zones (the two special zones are listed separately) with an average reduction of 12 \% from 7 \% to 6.13 \%. 
When considering the forecasting horizon of one day, at the aggregated level, the recency effect reduces the MAPE by 21 \% from 4.88 \% to 3.86 \%. At the bottom level, the MAPE values are reduced for twelve of the eighteen regular zones with an average reduction of 15 \% from 6.55 \% to 5.56 \%. \\

\textbf{\citet{taylor2007short}} evaluates several univariate methods on electricity demand from ten European countries ENTSOE public data (\hyperref[D14]{D14}).
A double seasonal ARMA model is introduced as a benchmark and a periodic Autoregressive model is implemented to take into account time-variation that cannot be captured by the seasonal ARMA. A double seasonal Holt-Winters exponential smoothing is implemented. It introduces an additional seasonal index with the correspondent extra smoothing equation in comparison with the seasonal Holt-Winters method. An intraday cycle exponential smoothing model is introduced as an alternative form of exponential smoothing for double seasonality. This formulation allows the intraday cycles to be represented by seasonal components and to update them at different rates using smoothing parameters. Finally, a method based on principal component analysis is implemented.
A static training methodology is used with the first twenty weeks of each load country as training set and the remaining ten weeks as testing set. 
The special days such as bank holidays are smoothed out by taking averages of observations from the corresponding period in the two adjacent weeks.
The MAPE are computed for each model and country for increasing horizons from a few hours to one day. Then, the scores are averaged over the ten countries. The double seasonal exponential smoothing performed the best from approximately 0.75 \% to 1.75 \%, followed by the principal component analysis, from approximately 0.9 \% to 2.05 \%, then the double seasonal ARMA from approximately 0.8 \% to 2.2 \% when increasing the horizon. Univariate methods are interesting for prediction up to about four or six hours ahead as the meteorological variables tend to change in a smooth fashion, which will be captured in the demand series itself. \\

\textbf{\citet{Taylor2008}} evaluates several univariate methods on demand from public National Grid minute-by-minute data (\hyperref[D13]{D13}). It is an extension of \citet{taylor2007short}. The study investigates also the lead time where a weather-based approach becomes superior to the univariate models.
A double seasonal ARMA model is introduced as a benchmark. An adapted version of the double seasonal Holt-Winters exponential smoothing is introduced to accommodate the two seasonal cycles in electricity load series. The double seasonal intraday cycle exponential smoothing implemented is an extension of the intraday cycle exponential smoothing developed in \citet{taylor2007short}. These models are compared to the weather-based approach used at National Grid \citet{taylor2003using}. This approach relies on weather-based regression models that are estimated independently on several cardinal points (such as evening peak, or strategically chosen fixed points) of the daily demand curve. Then, forecasts between cardinal points are built by fitting a curve with a ``profiling heuristic" procedure. It proceeds by judgmentally selecting a past load curve which is likely to be similar to the load profile for the next day. This method is in a way related to the similar day method mentioned in \citet{Weron2006,Weron2014} and \citet{Hong2016}. Finally, forecast combination is used by computing the simple average of the weather-based approach and the double seasonal exponential smoothing models.
The training methodology and the treatment of special days are the same as in \citet{taylor2007short}.
For increasing horizons from one to thirty minutes, the double seasonal exponential smoothing achieved the best MAPE values from 0.12 \% to 0.5 \%, the double seasonal intraday cycle exponential smoothing performed second, 0.45 \% to 0.52 \%, the double seasonal ARMA performed third, 0.12 \% to 0.75 \%, and the weather-based approach is the last one, 0.2 \% to 0.72 \%. This last approach is not competitive because it designed for lead times of at least several hours ahead to benefit from the weather forecasts.
For forecasting increasing horizon from one minute to one day, the combination of the weather-based approach and the double seasonal exponential smoothing achieved the best MAPE values from 0.15 \% to 1.2 \%, the weather-based approach ranked second, 0.15 \% to 1.3 \%, the double seasonal exponential smoothing ranked third, 0.15 \% to 1.4 \%, and the double seasonal intraday cycle exponential smoothing is the last, 0.15 \% to 1.8 \%.  \\

\textbf{\citet{Taylor2010}} implements triple seasonal univariate methods on demand from the public ENTSOE data (\hyperref[D14]{D14}). It is an extension of \citet{Taylor2008}.
Single, double and triple seasonal formulations of ARMA, Holt-Winter exponential smoothing and intraday cycle exponential smoothing are implemented. The triple seasonal formulations are extensions of double seasonal models considered in \citet{Taylor2008}. A single hidden layer feed-forward network is also implemented. A separate neural network model is built for each forecasting horizon. Finally, a combination of the triple seasonal ARMA and Holt-Winter exponential smoothing formulations is considered.
A static training methodology is used with the first five years as training set and the last remaining year as testing set. 
The treatment of special days is the same as in \citet{taylor2007short,Taylor2008}.
The horizons increase from thirty minutes to one day by step of thirty minutes. The triple seasonal ARMA and Holt-Winter exponential smoothing approaches outperformed the double ones. They have similar MAPE across all the forecasting horizons from 0.4 \% to 1.75 \%. The simple average combination of these two models led to a greatest accuracy from 0.35 \% to 1.6 \%. The feed-forward network was outperformed by the triple seasonal formulations with MAPE from 0.5 \% to 2 \%.

\subsection{Short-Term Load Forecasting}\label{annex:Review_STLF}

\textbf{\citet{Shepero2018}} use Gaussian process to compute probabilistic forecasts of residential consumption of an Australian distribution network (\hyperref[D15]{D15}). A log-normal process transforms the data before applying Gaussian process. This choice relies on the assumption that the electricity consumption of households is better represented by the log-normal distribution than a normal one \citet{Munkhammar2014}. This approach is compared to the conventional Gaussian process were the data are not transformed.
Twelve kernels are tested with two predictor options. The first option with hour of the days, thirty minutes and twenty-four hours time lags. The second option excludes the twenty-four hours time lag. Both Gaussian process approaches are implemented using the GPML package developed for MATLAB \citet{rasmussen2016gpml}. The first Mat\'ern kernel combined with the second predictor option performed the best in terms of PICP during the selection phase that consists in two years of data. These kernels are selected in the testing phase that consists in one year of data.
The training methodology is a blocked form of cross-validation. Four weeks of data are used for learning and one week to evaluate the error metrics. Then, a shift of one week is applied and the previous procedure is repeated.
Manual threshold is used to remove the anomalous measurements of the dataset, the detailed methodology is explained in \citet{Ratnam2017}.
The MAE and RMSE deterministic metrics and the PICP and PINAW probabilistic metrics are computed. Both approaches performed approximately the same MAE and RMSE values. The log-normal transformation produced sharper forecasts it in terms of PINAW. However, the PICP is deteriorated. The sharpness of the forecast varied throughout the day with the log-normal transformation. This behavior reflects the variation in the uncertainty of the forecasted value. This is not the case with the conventional approach as the sharpness of the forecast was almost constant throughout the day. Both methods were not able to capture the sudden sharp increments of the load. However, the temperature is not considered as predictor and may help both methods to capture this behavior. \\

\textbf{\citet{Ahmad2018}} use four machine learning techniques to forecast the electricity load of a water source heat pump on short and medium-term at a residential building level. The data are measured at site and composed of five minutes intervals of energy consumption and climate. 
Compact decision trees, fit $k$-nearest neighbor, linear and stepwise linear regressions models are implemented. The inputs of the models are climate variables (wind speed, wind direction, direct solar radiation on the surface), occupancy rate, previous week load, last twenty-four hours average load, previous day load, days of week, work days, and holidays.
The training methodology is static.  
A step of collecting, rescaling, harmonizing, cleaning, and formatting the data is done but the methodology is not discussed.
One week, two weeks, and one month ahead forecasting are studied. The MAE, MAPE, and RMSE are computed. For one week ahead forecasting the fit $k$-nearest neighbor achieved the best MAPE with 0.076 \%, followed by the compact decision trees with 0.115 \%, the stepwise linear regression with 0.340 \%, and the linear regression with 1.595 \%. For one month ahead forecasting, the compact decision trees achieved the best MAPE with 0.044 \%, followed by the fit $k$-nearest neighbor with 0.051 \%, the stepwise linear regression with 0.343 \%, and the linear regression with 0.776 \%. \\

\textbf{\citet{Goude2014}}, refer to \ref{32Goude2014} for description. \\

\textbf{\citet{Charlton2014}} propose a series of refinements of a linear regression model to forecast one week ahead hourly electricity demand of the Global Energy Forecasting Competition 2012 (\hyperref[D1]{D1}). This approach ranked first.
The starting linear regression model is a quadratic function of the temperature multiplied by a linear function of day number. Thus, it takes into account the temperature effect, the long-term trends, and the cross effects. Load and temporal hierarchical forecasting methodologies are used by splitting the data in $20 \times 24 \times 42 \times 2$ groups. There is one model for each of one the twenty zones, the twenty-four hours of the day, two seasons of the year, for weekdays, and for weekend. The temperature used for each zone comes from the weather station that fits the best the energy demand.
A first refinement consists in using a multiple weather station selection methodology to compute for each zone a linear combination of the five best weather stations that fits the best the energy demand. A second refinement consists in increasing the number of season from two to four, adding day of season terms (the day number within the season) to the linear function of calendar variables, and a special treatment for public holidays. A third refinement uses a multiplicative local correction factor computed by considering forecast and actual values. This method corrects any local systematic over or under estimation. An additive local correction has been tested but did not perform as well as the multiplicative correction. Finally, outliers removal is performed on each zone by discarding any days of data containing hourly loads smaller than 20 \% of the mean load of this zone. \\

\textbf{\citet{Lloyd2014}} use gradient boosting and Gaussian process to forecast one week ahead hourly electricity demand of the Global Energy Forecasting Competition 2012 (\hyperref[D1]{D1}). This approach ranked second.
A visual search for irregularities revealed two zones with anomalies. A special treatment is done for both of them.
The linear regression benchmark of~\citet{hong2010short} is used. The gradient boosting model is implemented with most of the default settings. The Gaussian process approach used three different kernels: one for backcasting, one for forecasting, and one for backcasting a specific load zone.
A gradient boosting model is implemented for each zone and the demand is modeled as a function of time of the day, time within the week, temperatures, and smoothed temperatures (all weather stations). The Gaussian process model combines the squared exponential and the periodic kernels. These kernels act upon temperature and smoothed temperature. The final prediction is a linear weighted average of the three previous models. The weights are chosen by hand, using the public test scores. \\

\textbf{\citet{Nedellec2014}} use a multi-scale approach to forecast one week ahead hourly electricity demand of the Global Energy Forecasting Competition 2012 (\hyperref[D1]{D1}). The final model combines three sub-models that correspond to the long, medium and short terms. This approach ranked third. The data of the long-term model are aggregated by month for each load zone and weather station. The medium-term model is fitted on the detrended data, and the short-term model on its residuals.
The long-term model captures the low-frequency variations such as trends and economic effects, the medium-term model the daily to weekly effects (incorporating all of the meteorological effects such as temperature and the calendar effects). Finally, the short-term model takes into account everything that could not be captured on a large temporal scale but could be captured locally in time (close to the date of the prediction). There is a model per load zone and a medium-term model fitted per instant of the day. The best weather station per load zone is selected with a cross-validation strategy.
The long and medium terms approaches are general additive models. The short-term approach is a random forest. The inputs are the monthly load and temperature, the day type, the time of the year, and a smoothed temperature. 
The short-term model provides an average gain per load zone of 5 \% of the RMSE. \\

\textbf{\citet{bentaieb2013}} use a gradient boosting approach to forecast one week ahead hourly electricity demand of the Global Energy Forecasting Competition 2012 (\hyperref[D1]{D1}). This approach allows to take advantage of the good performance and the automatic variable selection. In addition, the boosting algorithm and the penalized regression splines provide a smooth estimation of the demand. This approach ranked fourth.
Forty-three potential predictors are identified. The demand is modeled with calendar effects (time of the year, day of week, holidays, etc.), past demand, current and past temperatures. There is a model per load zone and for each hour of the day. One weather station is selected per zone. Outliers are identified and replaced following a model-based with a fixed threshold methodology.
The demand is forecasted by using general additive models. Gradient boosting is used to estimate each model including variable selection during the fitting process. \\

\textbf{\citet{hong2010short}} propose a formal study of short-term electric load forecasting on a medium US utility (\hyperref[D2]{D2}). Three forecasting techniques are studied with different amount of emphasis. A systematic approach to investigate short-term load forecasting and to improve the forecasting accuracy is proposed. It relies on the selection of relevant features including the calendar variables (month, week day and holiday effects), the cross effects (temperature with calendar variables and between calendar variables), the recency effects of load and temperature, and Gross State Product as an extension for medium and long terms forecasting. 
Several linear regression models with consecutive levels of refinement are implemented starting from a benchmark model named General Linear Model based Load Forecaster - Benchmark 7 (GLMLF-B7). The GLMLF-B7 models the consumption as a cubic function of temperature and calendar variables, including cross effects. A first refinement consists in adding the temperature recency effect by including lagged temperature variables. A second refinement captures the weekend and holiday effects by grouping the days of week and modeling the special days. Finally, an exponentially weighted least square approach is implemented by assigning higher weights to the recent observations than the older ones. 
A Possibilistic Linear Model Based Load Forecasters named (PLMLF-B7), that is a fuzzy regression model, is implemented with the same predictors than GLMLF-B7.
Several single-output feed-forward neural networks with consecutive refinements are implemented. The first one is named ANN Based Load Forecaster - Benchmark (ANNLF-BS) and is similar to GLMLF-B7. Then, several refinements, similar to the ones performed on the linear regression models, are implemented. In addition, temporal hierarchical forecasting is considered by testing several series of multiple feed-forward neural networks in parallel. A series of models for each hour of the days, a series for each day of the week, and a series for each month of the year. The study demonstrated that forecasting each hour, day of the week, or month separately does not help to improve the forecasting accuracy on this dataset.
A sliding window methodology is implemented. Three years of data are used as learning set. Then, depending on the horizon, it is shifted by one hour, one day, one week, two weeks, or one year.
The results indicate that linear regression models can be more accurate than feed-forward neural networks and fuzzy regression models given the same amount of input information. In addition, the linear regression benchmark can be used for medium and long terms forecasting by including some correction to taken into account the long-term effects.
The final refined linear regression short-term model has been deployed in a US utility for production use and the benchmark model (GLMLF-B7) has been implemented as a base model in the commercial software package SAS Energy Forecasting\footnote{https://www.sas.com/en\_us/home.html}.
The MAPE values of GLMLF-B7 are 4.96 \%, 4.98 \%, 5.04 \%, 5.06 \%, and 5.20 with horizons of one hour, one day, one week, two weeks, and one year. The MAPE values of the final version deployed in a US utility are 2.97 \%, 3.06 \%, 3.17 \%,  3.23 \%, and 3.44 \% for the same horizons. 
The MAPE values of PLMLF-B7 and GLMLF-B7 for a forecasting horizon of one year are 7.56 \% and 5.20 \%. 
The MAPE of ANNLF-BS and a refined version named ANNLF-HTS for a forecasting horizon of one year are 6.51 \% and 4.51 \%. \\ 

\textbf{\citet{HongWang2014}} propose a fuzzy interaction regression approach to forecast one day ahead hourly load of ISO New England (\hyperref[D10]{D10}). Four tips on fuzzy interaction regression for load forecasting are discussed. In addition, three critical comments to a notable but questionable paper of the field on its parameters estimation, forecasting results and conclusions are provided.
A sliding window methodology with a training set of two years of hourly load and temperature is used. The forecasts are computed on a daily rolling basis on the last third year of the dataset.  
The fuzzy interaction regression approach is implemented in the earliest possibilistic regression framework of \citet{asai1982linear}. Three variants of models are considered. A generic model named M1 without calendar variables, only considering a third order polynomial of temperature including several temperature lags. M2 is a refined generic model with the calendar variables: hour, weekday, month. M4 is a refined version of M2 with the cross effects: temperature and hour of the day, temperature and month of the year, hour of the day, and day of the week. These models are compared to a linear regression model named M3 that has the same predictor variables than M2.
Overall, the most refined fuzzy interaction regression model M4 demonstrated superior accuracy in terms of MAPE over the other models.
The MAPE of the models are computed for the hourly load, the daily peak, the daily energy, and the annual and winter peak day. The values for the hourly load for M1 to M4 models are 14.2 \%, 15.16 \%, 4.63 \%, and 3.68 \% . \\

\textbf{\citet{Qahtani2013}} propose a multivariate $k$-nearest neighbor approach to forecast on day ahead hourly UK electricity demand.
The multivariate approach is compared to a univariate $k$-nearest neighbor and a statistical benchmark model. The univariate approach consider only past values of the load as the multivariate approach includes also calendar variables: regular working days, weekends, and bank holidays. Both $k$-nearest neighbor approaches consider the Euclidean distance function. A grid search is conducted to set the metaparameters of the algorithm: the number of vectors to be compared to the target, $k$, and $m$ their dimension. Each vector is composed of $m$ consecutive observations.
The training methodology is static with one complete year from the National Grid data (\hyperref[D13]{D13}) used as training set. The models are assessed in predicting all days of the following year.
The proposed $k$-nearest neighbor outperformed all benchmark models. The MAPE of the multivariate and univariate approaches are 1.81 \% and 2.38 \%. \\ 

\textbf{\citet{Lopez2018}} propose an on-line model to forecast in real-time nine days ahead hourly demand for the Spanish transport system operator REE (\hyperref[D5]{D5}). This is a hybrid load-forecasting based on an Autoregressive model and neural networks techniques.
The training methodology is static. Several training periods and retraining frequencies are tested. More than three years of data and retraining more frequently than once a year did not provide any further improvement. Data cleansing is performed by identifying abnormalities. The data are compared to an expected range based on forecasts and past load.
Four sub-models are used to build the national forecast. Two models forecast directly the demand at the national level: an Autoregressive and a neural network models. In addition, two other Autoregressive and neural network models forecast at the regional level (eighteen zones). Then, the forecasts are aggregated to build a forecast at the national level. Finally, the final forecast at the national level is a linear combination of these four models. The models take as inputs weather variables, calendar information such as special days, and temperature lags. The most relevant weather stations are selected for each region among fifty-nine stations scattered across Spain. Cold and hot degree days are used to pre-process the data to address the nonlinear relationship between the load and temperature.
The RMSE and MAPE values for the Autoregressive model that forecasts directly at the national level are 3.66 \% and 3.33 \%. In comparison with 2.95 \% and 2.50 \% for the neural network. For the Autoregressive model that forecasts at the regional level the scores are 2.29 \% and 1.97 \%. In comparison with 3.09 \% and 2.69 \% for the neural network. Finally, the scores are 1.83 \% and 1.56 \% for the final forecast.

\subsection{Medium-Term Load Forecasting}\label{annex:Review_MTLF}

\textbf{\citet{Xie2015LTLF}} propose linear regression and survival analysis to forecast the long-term consumption of a retailer portfolio. The methodology consists in three steps: forecasting the load per customer, forecasting the tenured customers, and multiplying both forecasts to compute the prediction of the tenured demand. This methodology has been validated through a field implementation at a fast growing U.S. retailer. 
The training methodology is static and the dataset (\hyperref[D8]{D8}) includes for each customer the hourly temperature history of the associated weather stations.
The variable selection methodology of \citet{hong2010short} is used to customize a linear regression benchmark as starting model. The set of predictors is composed of a trend, a third order polynomial of the temperature, calendar variables with month, day and hour, and seven cross effects between temperature and calendar variables. Then, the customized versions are built by adding the recency temperature, weekend, and holidays effects. This model is used to forecast the load per customer. Then, the survival analysis is used to model and forecast the tenured customers.
The MAPE values of the load per customer forecast for hourly load, daily energy and monthly energy are 11.56 \%, 10.03 \%, and 7.75 \%. Two tariff plans are proposed by the retailer to the customers: a fixed plan with a constant purchasing price and a variable plan with a fluctuating purchasing price based on market conditions. Each plan requires its own forecast of the tenured customers. The MAPE values of the final forecast for the fix plan are 10.24 \%, 8.99 \%, and 7.45 \%. In comparison with 10.56 \%, 9.51 \%, and 7.96 \% for the variable plan. The survival analysis methodology enables to divide by two the MAPE values of the final forecast in comparison with a common method used in the industry for the fix plan. \\

\textbf{\citet{Xie2015}} propose a combination of several forecasting techniques to address the NPower Forecasting Challenge 2015. The dataset (\hyperref[D4]{D4}) is divided into three parts for each round of the competition. The training set for parameter estimation, the validation set for model selection, and the testing set. The weather and calendar data for the forecasting period are provided at each round.
Four forecasting techniques are considered. A linear regression model that is similar to the benchmark of \citet{hong2010short}. An ARIMA model is implemented and a trial-and-error method is used to identify the order of the autoregressive model, the degree of the differencing, and the order of the moving average term. A feed-forward neural network with calendar and weather variables, similar to the linear regression model, is used. Finally, a random forest model with five hundred trees is implemented with weather variables and for the last round calendar variables. The final forecast is the average of the four individual forecasts at each round.
A forward and a backward selection strategies are used to select the relevant input variables of the linear regression model. They are both similar in terms of methodology to pre-pruning and post-pruning for decision trees. The first strategy consists in adding variables one by one to a simple model such as a linear regression benchmark. If an additional variable contributes to decrease the MAPE it is kept. The backward selection strategy consists in eliminating variable one by one from a more complex model. If a removed variable does not result in increasing the MAPE it is definitely eliminated. 
This methodology ranked third. The average MAPE over the three rounds of the final forecast is 2.40 \% and better than any of the individual model. The best linear regression model achieved  2.47 \%, the neural network 3.27 \%, the random forest 3.87 \%, and the ARIMA 8.23 \%.\\


\textbf{\citet{Goude2014}}\label{32Goude2014} propose a methodology to forecast, both at short and medium terms, the consumption of 2200 substations of the French distribution network ENEDIS (\hyperref[D3]{D3}). It is a semiparametric approach based on generalized additive models.
The training methodology is static with a learning set composed of the first five years and the testing set of the last year. The data cleansing consisted in excluding 360 time series as they contain too many outliers to estimate the model.
The forecasting horizon is decomposed in short and medium terms with dedicated general additive models. The medium-term model takes as inputs lags of the load, current and smooth temperatures, day type (including bank holidays), and time of the year. The short-term model is derived from this model by adding a lag load effect. Then, a refined version of the medium-term model, named medium-term detrending model, is built by modeling the trend. It consists in detrending the data at a monthly scale and fitting the model. Finally, the forecasts are computed by aggregating the detrended forecasts and the estimated monthly trend. 
Load and temporal hierarchical forecasting methodologies are considered. There is a model fitted per substation and per instant of the day. A weather station is affected to each substation by a meteorologist based on climate properties criteria.
This approach is able to forecast more than 2 000 electricity consumption series by capturing automatically the variety of each substation. The short-term model achieved a median MAPE of 5 \% for one day ahead forecasting. The medium-term model and the detrending extension achieved median MAPE values of 8 \% and 6 \% for one year ahead forecasting. \\

\textbf{\citet{Gaillard2016}} use a concatenation of a short and medium-term models to forecast one month ahead hourly electricity probabilistic demand of the Global Energy Forecasting Competition 2014 (\hyperref[D6]{D6}). The probabilistic forecasts are computed by producing temperature scenarios. This approach ranked first.
The forecasting horizon is decomposed in two parts with a dedicated quantile generalized additive model. The short-term horizon is two days and the medium-term model forecasts from two days to one month ahead. The final forecast is the concatenation of both models.
The inputs of the models are an exponential smoothed of the temperature, a recent lag of temperature for the short-term model, the time of the year, and the day type. The temperature comes a from virtual weather station. It is a simple average of four weather stations that are selected by computing generalized cross-validation scores. Data segmentation is performed by dividing the dataset into twenty-four independent time series, one per hour of the day. Then, twenty-four separate models are fitted.
The PLF values are between 4 and 11.    \\

\textbf{\citet{Dordonnat2016}} derive from a semiparametric regression model for point load forecasting a strategy to forecast one month ahead hourly electricity probabilistic demand of the Global Energy Forecasting Competition 2014 (\hyperref[D6]{D6}). This approach ranked second.
A temperature based deterministic load model is built with a generalized additive model. A multivariate simulation model for the twenty-five weather stations is implemented to produce temperature sample paths. Then, they are used as inputs of the deterministic load model to compute load sample paths. The final probabilistic forecast composed of quantiles is derived from final load paths. 
The generalized additive model takes as inputs the temperature, an exponential smoothed temperature, and the day type. The temperature comes from virtual weather station. It is an average temperature of three weather stations selected by an exponentially weighted average algorithm.
The MAPE metric is used to select the best deterministic load model. The MAPE values are between 8.74 \% and 11.83 \% depending on the features. However, the most accurate deterministic model in terms of MAPE will not necessarily lead to the best probabilistic forecast when using temperatures from simulations. The PLF of the probabilistic models derived are between 7.37 to 8.37. \\

\textbf{\citet{xie2016}} propose an integrated solution with forecast combination and residual simulation to forecast one month ahead hourly electricity probabilistic demand of the Global Energy Forecasting Competition 2014 (\hyperref[D6]{D6}). The final solution is composed of three parts: pre-processing, forecasting, and post-processing. This approach ranked third.
The pre-processing part includes data cleansing using the linear regression benchmark of \citet{hong2010short} and the temperature station selection framework of \citet{hong2015weather} by averaging the temperature of the top eleven weather stations.
The forecasting part is subdivided into three steps. The first one involves the development of a linear regression deterministic model. The variable selection methodology of \citet{hong2010short} is used to customize the model that produces a point load forecast. The residuals are fed to the second stage where they are modeled. Then, the residual forecast is computed by using four forecasting techniques: exponential smoothing, ARIMA, a feed-forward neural network, and an unobserved components model. It leads to four second-stage forecasts by adding the four second-stage residual forecasts to the first-stage forecasts. The second step consists in averaging the four second-stage forecasts to compute the point forecast combination. Then, the probabilistic forecasts are computed by using temperature scenarios.
Finally, the post-processing part simulates the residuals of the selected point forecasting models in order to improve the probabilistic forecasts.
This integrated approach is compared to twelve models that are divided into two groups: with and without residual simulation. Each group is composed of six underlying models: five from first-stage models and one from the forecast combination. On average, residual simulation in the post-processing step helped to improve the forecasts. The PLF of the final solution are between 3.360 and 11.867.  \\

\textbf{\citet{Haben2016}} propose a hybrid model of kernel density estimation and quantile regression to forecast one month ahead hourly electricity probabilistic demand of the Global Energy Forecasting Competition 2014 (\hyperref[D6]{D6}). This approach ranked fourth.
The hybrid model is a combination of five models each one forecasting on a dedicated time period. Period one is the first day, period two the rest of the first week, period three the second week, period four the third week, and period five the rest of the month. 
These models are mainly based on kernel density estimation and quantile regression. Three based kernel techniques are implemented. A kernel density estimation with a time decay parameter to give a higher weight to more recent observations. A conditional kernel density estimation on the period of the week named CKD-W. It provides a higher weight to observations from similar hourly periods of the week. A conditional kernel density estimation on the temperature named CKD-T. The variables of the kernel based techniques are the mean hourly temperature from the twenty-five weather stations. As the temperature forecasts are inaccurate beyond a few days, this method was only implemented for the first days of the horizon to forecast. Finally, the quantile regression technique is a simple linear function created separately for each hour of the day based on only the trend and seasonal terms. 
Each of these forecasting techniques performed differently depending on the forecast horizons. Several combinations are tested and two main mixed new forecasters are derived from the main models. The first mixture is only composed of the CKD-W model except for using the CKD-T model on the first day. The second one is the CKD-T for the first day, then the CKD-W from the second to the seventh day, and quantile regression from the eighth day up to the end of the month. Finally, the final hybrid model is built as a combination of five models each one forecasting on a dedicated time period. The CKD-T model forecasts over the first period. The models of the other periods are a weighted average of quantile regression and CKD-W models.
The final hybrid model had the best accuracy in comparison with the two mixed forecasters, and each of the main model taken independently. \\

\textbf{\citet{Chen2001}} propose a support vector regression model to forecast one month ahead daily electricity demand of the EUNITE Competition 2001 (\hyperref[D16]{D16}). This approach ranked first.
The feature selection process considers calendars (weekdays, weekends, and holidays), normalized temperature, and the past seven daily maximum loads variables. Data segmentation is used by only considering the winter data segment for training.
Data segmentation seemed to enhance the model performance in terms of MAPE. Surprisingly models built with temperature did not achieve the best results. However, due to the limited amount of data it was very difficult to achieve accurate forecasts of the temperature. This study demonstrates that a conservative approach using only available correct information is recommended in this case. The MAPE values are between 1.95 \% and 2.97 \% depending on the inputs (with or without temperature) and the length of the winter segment considered. The best MAPE is achieved without considering temperature. \\

\textbf{\citet{Wang2016}} refer to \ref{12Wang2016}. \\

\textbf{\citet{Ziel2016}} propose a methodology based on Least Absolute Shrinkage and Selection Operator (LASSO) estimation to forecast one month ahead hourly electricity probabilistic demand of the Global Energy Forecasting Competition 2014 (\hyperref[D6]{D6}). This approach is also applied to an extension of this competition (\hyperref[DD7]{D7}) where it ranked second. The LASSO methodology has the properties of automatically selecting variables making a useful tool for feature selection.
A bivariate time-varying threshold Autoregressive model for the hourly load and temperature is implemented. The modeling process has three crucial components. The choice of the thresholds sets, the lag sets, and the time-varying structure of the coefficients. Eight groups of variables are considered: hourly impacts on the seasonal daily pattern, hourly impacts on the seasonal weekly pattern, daily impacts on the seasonal annual pattern, smooth annual impacts, long-term trend effects, fixed date public holidays effects, varying date public holidays effects, and interaction effects between the first and fourth groups. LASSO methodology automatically gives low or zero values to the parameters of less important variables. Given the estimated model, a residual-based bootstrap is used to simulate future scenario sample paths (10 000). For the Global Energy Forecasting Competition 2014, the temperature is the average of two weather stations.
The approach is compared to two benchmarks: the linear regression model of \citet{hong2010short} and the extension with recency effect of \citet{Wang2016}. The PLF is 7.44 for the Global Energy Forecasting Competition 2014 and 54.69 for its extension. The PLF reductions are of 6.4 \% and 7.6 \% in comparison with both benchmarks for the Global Energy Forecasting Competition 2014. The reductions are of 11.9 \% and 15.6 \% for the extended version of the competition. \\

\textbf{\citet{Xie2017}} investigate the normality assumption in residual simulation for probabilistic load forecasting on two cases studies: the North Carolina Electric Membership Corporation (\hyperref[D9]{D9}) and the Global Energy Forecasting Competition 2014 (\hyperref[D6]{D6}). The goal is to understand whether applying the normality assumption in residual simulation helps to improve the quality of probabilistic load forecasts. 
Three linear regression models and three feed-forward neural networks are implemented with an increasing number of input features. Both techniques are applied to both case studies. The starting linear regression model includes a macroeconomic trend, the Gross State Product, and a third-order polynomial of temperature. The second one is the linear regression benchmark of \citet{hong2010short} that includes several calendar variables and cross effects with temperature. The third one is a refined version of both first models and includes a macroeconomic trend, the recency effect modeled by lags of temperature, the weekend, and holiday effects. The relevant variables of this model are selected following the variable selection framework of \citet{hong2010short}. Concerning the three feed-forward neural networks, they have the same inputs variables as the three linear regression models with the high order terms and cross effects removed. 
The probabilistic forecasts are computed by generating weather scenarios. They are fed to the point load forecast models that produce several load forecast paths. Then, the original probabilistic forecast is post-processed by adding simulated residuals generated from a normal distribution.
Overall, modeling residuals with a normal distribution helps to improve the accuracy when the forecasters do not have enough resources to build comprehensive underlying models. However, the improvement is diminishing with the refinement of the underlying model. A very comprehensive underlying model will not benefit from residual simulation based on the normality assumption. The differences between the PLF without and with residual simulation are 33.2, 0.9, and 0.2 for the three linear regression models on the North Carolina Electric Membership Corporation case study. The same trend is observed when considering the Global Energy Forecasting Competition 2014 case study with all the linear regression models and feed-forward neural networks implemented. This study demonstrates the importance of sharpening the underlying model before post-processing the residuals.\\

\textbf{\citet{hyndman2015monash}} propose a methodology to forecast the probabilistic distribution of annual, seasonal and weekly peak electricity demand and energy consumption for various regions of Australia. The model is capable of forecasting both the half hourly short demand and the probability distributions. It relies on demand and temperature data that are available on a half-hourly basis AEMO data (\hyperref[D12]{D12}), while the economic and demographic data are only available on a seasonal basis.
The forecasting horizon is decomposed in a short and long terms. The short-term model is based on half hourly demand and weather information. A model is fitted for each half-hourly period. The long-term model is based on the seasonal demographic, economic variables, and degree days (all linear terms). 
A semiparametric additive model estimates the relationships between demand and the driver variables, including temperatures, calendar effects and some demographic and economic variables. The average and difference temperature of two weather stations are considered as the temperatures at the two locations are probably highly correlated. The dataset is split into a morning, afternoon and evening subsets. The best model is selected for each subset separately.
Then, the forecast distributions are derived from the model using a mixture of temperature and residual simulations, and future assumed demographic and economic scenarios.
Ex-ante and ex-post forecasts are computed. Ex-ante forecasts are based only on information available at the time of the forecast, whereas ex-post forecasts use information beyond the time at which the forecast is made. The difference between both forecasts provide a measure of the performance of the model. Unfortunately, there is no quantitative metric used in this study. But as it is pointed out, it is difficult to evaluate distributional accuracy on annual maxima because there is only one actual value per year. Thus, the weekly maxima are used to evaluate the distributions and the twenty-six actual weekly maximum demand values fall within the region predicted from the ex-ante forecast distribution.

\subsection{Long-Term Load Forecasting}\label{annex:Review_LTLF}

\textbf{\citet{Hong2008}} propose a long-term spatial load forecasting using human-machine co-construct intelligence framework to address the Madison case study (\hyperref[D17]{D17}). 
The forecasting module consists mainly in two parts. The first one is the bottom-up module that aggregates historical load for each small area in each level to compute the S-curve parameters. The second one is the top-down module takes the bottom-up module as inputs, and allocates the utility’s system forecast from the top to the bottom level. The forecasting module takes as inputs the outputs of the weather normalization, the horizon year load, and neighborhood modules. Finally, a human expert (human-machine co-construct intelligence) is integrated to provide heuristics and insights to correct or confirm the results from this automated computer program. 
The S-curve modeling is typical of a small area, distribution-level load growth, which has three distinct phases: a dormant period with no load or growth before development, a rapid growth in the small locations under construction, and a saturated period with slow growth in the small locations being fully developed. At each level S-curve is used to fit the corresponding data from each small locations. The bottom-up module aggregates all the small locations to obtain the load of the sub-region in the upper level.
The results are presented both in data and map formats. Unfortunately, there is no error metric to assess the results. However, the proposed method has been applied to several utilities and has received satisfactory results.\\

\textbf{\citet{HongXie2014}} propose a modern approach that takes advantage of hourly information to forecast long-term probabilistic demand on the North Carolina Electric Membership Corporation case study (\hyperref[D9]{D9}). This approach modernizes three key elements: the predictive modeling, the scenario analysis, and the weather normalization. 
The predictive modeling consists in implementing several deterministic linear regression models and to select the best ones. Four starting models are implemented. One for monthly energy forecasting, another one for peak forecasting, the benchmark of \citet{hong2010short} and a group of customized short-term models that are derived from the model selection methodology developed by \citet{hong2010short}. The short-term models are extended to long-term forecasting by adding the Gross State Product macroeconomic indicator. 
The scenario analysis consists in computing the probabilistic forecasts by using thirty weather scenarios and three macroeconomic scenarios, resulting in ninety cross scenarios in total. 
Finally, the weather normalization consists in using weather normalization to estimate the normalized load profile.
A sliding window methodology is used.
The most accurate extension of the short-term load models is achieved by replacing the trend by the Gross State Product. It achieved a MAPE of 4.7 \% for one year ahead forecasting and a window of three years. Then, the impact of the window length is assessed. Two years of data offers the lowest MAPE with 4.2 \%. Finally, all the linear regression models are compared in terms of MAPE for the annual energy, the annual peak, the monthly energy, the monthly peak, and the hourly load. The short-term models extended with the Gross State Product performed the best MAPE values. On monthly energy and peak forecasting, this approach reduces the MAPE by over 45 \% in comparison with the other approaches. The results of the probabilistic forecasting and weather normalization are depicted by Figures with the 10\textsuperscript{th}, median and 90\textsuperscript{th} percentiles. \\

\textbf{\citet{HongShahidehpour2015}} propose a review of load forecasting topics for states, planning coordinators, and three case studies for the Eastern Interconnection States' Planning Council: North Carolina Electric Membership Corporation (\hyperref[D9]{D9}), ISO New England (\hyperref[D10]{D10}), and Exelon Corporation (\hyperref[D11]{D11}).
The North Carolina Electric Membership Corporation is divided into three supply areas, ISO New England has eight zones in six states, of which Massachusetts has three zones, and Exelon Corporation has three operating companies in central Maryland, southeastern Pennsylvania, and northern Illinois.
The case studies are addressed with a linear regression technique similar to \citet{HongXie2014} and customized for long-term forecasting. 
The main effects are the temperature with a third order polynomial, the calendar variables, and a macro economic indicator. The cross effects are several cross variables of temperature and calendar variables.
Hierarchical load forecasting is implemented by forecasting at the zonal and aggregated levels of each case study. The weather station selection process of \citet{hong2015weather} is used for the North Carolina Electric Membership Corporation case study and the combination of weather stations varies from one supply area to another.
The results are depicted on Figures. For the North Carolina Electric Membership Corporation case study, the ex-ante probabilistic and point forecasting monthly peak and energy results for each of the three supply areas are provided in comparison with the actual load for the years from 2009 to 2014. 
For the ISO New England case study, the ex-ante probabilistic forecasting monthly peak and energy results for each zone are provided in comparison with the actual values for the years 2014 and 2015.
For the Exelon Corporation case study, the ex-ante probabilistic and point forecasting monthly peak and energy results for each of the three operating companies are provided in comparison with the actual load for the years from 2011 to 2013.

\section{Datasets}\label{annex:Datasets} 

Tables \ref{tab:LFdatareview1} and \ref{tab:LFdatareview2} list the datasets used in the articles reviewed and provides key information about the dataset composition (load, temperature, etc.), the system description (one or several zones, typical load values, etc.) and the dataset access (free or private).

\begin{table*}[htb]
	\renewcommand\thetable{9.1}	
	\renewcommand{\arraystretch}{1.5}
	\caption{Datasets used into the articles reviewed, first part. \label{tab:LFdatareview1}}	
	\begin{tabular}{m{0.5cm}m{3cm}||m{3.5cm}m{6.5cm}m{1.5cm}}
		Id    &  Dataset  & Data  & System description   & Access \\  \hline \hline
		
		D1\label{D1}  & Global Energy Forecasting Competition 2012 load track & Four years and a half of hourly load and temperature. & US utility with twenty zones, each from a few MW to 200 MW. The total system load is about 1 800 MW. Eleven weather stations. Data available through \citet{hong2014global}.  & \cmark \\
		
		D2\label{D2} & Medium US utility & Nine years of hourly load and temperature. & Medium US utility with a load from 100 MW to 1 GW.   & \xmark \\
		
		D3\label{D3} &  ENEDIS & Six years of load and temperature. & 2 260 substations at the frontier between the high voltage grid and the distribution network and sixty-three weather stations in France. & \xmark  \\	
		
		D4\label{D4} & RWE NPower Forecasting Challenge 2015  & Two years of daily load, weather and calendar. & Daily consumption of RWE NPower customers from 50 MWh to 90 MWh . & \xmark  \\	
		
		D5\label{D5} & Spanish Transport System Operator &Ten years of hourly load.  & Load at the aggregated level from 10 GW to 50 GW available on ENTSOE website, fifty-nine weather stations.  & \cmark \\	
		
		D6\label{D6}  & Global Energy Forecasting Competition 2014-L load track & Eleven years of weather and five years of load data. &One load zone with values from 60 MW to 310 MW and twenty-five weather stations. Data available through \citet{hong2016probabilistic}. & \cmark  \\
		
		D7\label{D7} & Global Energy Forecasting Competition 2014-E load track & Six years of hourly temperature and four years of hourly load. & One load zone with values from 1.8 MW to 5.5 GW and one weather station. Data available through \citet{Hong2016}.   & \cmark  \\
		
		D8\label{D8}  & Retail electricity provider data & Three years of hourly load and temperature data.  & One load zone of a retail electricity provider with values from 0 to 19 MW.  & \xmark \\		
		
		D9\label{D9} & North Carolina Electric Membership Corporation & Ten years load, economy and thirty years of weather data. & Load at system level with values from 1 to 5 GW. Three areas with load values from 200 to 1 000 MW. & \xmark \\	
		
	\end{tabular}
\end{table*}

\begin{table*}[htb]	
	\renewcommand\thetable{9.2}
	\renewcommand{\arraystretch}{1.5}
	\caption{Datasets used into the articles reviewed, second part. \label{tab:LFdatareview2}}	
	\begin{tabular}{m{0.5cm}m{3cm}||m{3.5cm}m{6.5cm}m{1.5cm}}
		Id    &  Dataset  & Data  & System description   & Access \\  \hline \hline
		
		D10\label{D10}  & ISO New England & Years of load, weather and economical data. & System level (eight zones): 15 - 30 GW. Zonal level: 700 - 8 000 MW. Data available on ISO NE website. & \cmark \\	
		
		D11\label{D11}  & Exelon & Years of load, weather and economical data. & Exelon: three operating companies. System level: 10 - 45 GW. Company level: 3 - 25 GW. Data available on PJM website. & \cmark \\		
		
		D12\label{D12} & AEMO & Half-hourly load profiles since 2002.& Victoria, New South Wales, South Australia, Tasmania, Queensland Australian regions. Aggregated level: 20 - 30 GW. Regional level: MW - GW. Data available on AEMO website. & \cmark \\	
		
		D13\label{D13} & National Grid & Half-hourly load profiles since 2005. & Load from 20 - 50 GW.  Data available on National Grid website.   &  \cmark\\	
		
		D14\label{D14}  & ENTSOE & Historic electricity demand of European countries. & Load country from GW to 50 GW. Data available on ENTSOE website. & \cmark \\	
		
		D15\label{D15}  & Ausgrid residential dataset &Three years of half-hourly load and generation. & Load and rooftop PV generation for 300 de-identified of residential Australians. Data available on Ausgrid website.  & \cmark \\		
		
		D16\label{D16}  & EUNITE Competition 2001 & Two years of half-hourly load, four years of average daily temperature.  & Load values from 400 to 800 MW. Data available on EUNITE 2001 website. & \cmark \\	
		
		D17\label{D17}  & Madison data & Electric load history, current and future land use information.  & Madison in Wisconsin divided into hundreds of zones from a few kW to hundreds of kW electricity demand. & \xmark \\	
		
	\end{tabular}
\end{table*}

\newpage

\section{Data cleansing techniques}\label{annex:DCT}

The data cleansing techniques are poorly discussed in the literature. Datasets from day life applications often require post-processing. The way of doing these corrections have an impact on the forecasting results. Table \ref{tab:DCT} presents the basic principles of the data cleansing techniques implemented in the articles reviewed.

\begin{table*}[!htb]
	\renewcommand\thetable{10}
	\renewcommand{\arraystretch}{1}
	\caption{Data cleansing techniques.} 
	\label{tab:DCT}
	\begin{tabular}{m{2cm}||m{3cm}m{9.5cm}}
		Technique   & Author  & Principles  \\ \hline   \hline
		
		Manual threshold & \citet{Ratnam2017} \citet{VanderMeerShepero2018} \citet{Shepero2018}  & Loads with a maximum value $<$ 6 W, PV solar generation with a maximum value $<$ 0.06 kW, a daily generation $<$ 0.325 kWh and an early morning (before 5 am) generation $>$ 0.02 kWh are anomalies and removed from the dataset. \\ \hline
		
		Naive method & \citet{LUO2018}  & Load values outside the interval $[\sigma - k* \mu , \sigma + k* \mu ]$ are treated as anomalies, where $k$ is a threshold fixed manually, $\sigma$ and $\mu$ are the mean and standard deviation of the load. \\ \hline
		
		Model-based with an adaptive threshold & \citet{LUO2018}   & A dynamic linear regression-based anomaly detection method with an adaptive anomaly threshold. \\  \hline
		
		Model-based with a fixed threshold & \citet{xie2016} \newline \newline \citet{bentaieb2013} & Load values where the APE is $>$ 50 \% in comparison with a linear regression model predictions are outliers and replaced by the predicted values. 
		\newline $y_{t}$ is an outlier if $ [y_{t} - \tilde{y_{t}}] > \text{median}  [y_{t} - \tilde{y_{t}}]  + k * \text{MAD} $ where $\tilde{y_{t}}$ is the Loess fit of $y_{t}$, MAD the mean absolute deviation and $k$ chosen so the probability of an outlier is $0.002$ under a normal distribution.\\	\hline
		
		Seasonal naive method & \citet{Charlton2014} \citet{LUO2018}  & The load values at hour $h$ outside the interval $[\sigma_{h} - k* \mu_{h} , \sigma_{h} + k* \mu_{h} ] $ are treated as anomalies, where $k$ is a threshold fixed manually, $\sigma_{h}$ and $\mu_{h}$ are the mean and standard deviation of the load at hour $h$. \\ \hline
		
		Smoothing specific values & \citet{taylor2007short} \citet{Taylor2008} \citet{Taylor2010} &  Load values of special days such as bank holidays are smoothed out with load averages from the corresponding period in the two adjacent weeks. \\ \hline
		
		Visual analysis & \citet{Lloyd2014}  & Graphical representations of data with box plots, etc. 
	\end{tabular}
\end{table*}

\newpage

\section{Error measurement metrics}\label{annex:EM} 

\citet{gneiting2014probabilistic}, \citet{Hong2016} and \citet{van2017review} did a review of the performance metrics. Table \ref{tab:EMdeterministic} and \ref{tab:EMproba} list the deterministic and probabilistic metrics implemented in the articles reviewed.

\begin{table*}[!htb]
	\renewcommand\thetable{11.1}
	\renewcommand{\arraystretch}{1}
	\caption{Deterministic forecasting error metrics. } 
	\label{tab:EMdeterministic}
	\begin{tabular}{m{1.1cm}||m{3cm}m{10.5cm}}
		EM   & Author &  Principles  \\  \hline  \hline
		MAE   & \citet{Willmott2005} \citet{van2017review} & Mean Absolute Error is useful to compare several forecasts of the same time series. However, it is scale dependent and cannot be used to compare forecasts of different time series.  \\ \hline 
		MAPE  & \citet{van2017review} \citet{hyndman2006another}  & The Mean Absolute Percentage Error is a normalized MAE. Common choices for the normalization factor are the mean or the range (the maximum minus the minimum) of the data. It is scale invariant and allows to compare different time series.  \\ \hline 
		MASE  & \citet{hyndman2006another} & The Mean Absolute Scaled Error is MAE scaled with the in-sample MAE of the naive method (random walk). It is scale invariant, symmetric thus penalizes positive and negative forecast errors equally and penalizes errors in large forecasts and small forecasts equally.  It is easily interpretable, as values greater than one indicate that forecasts from the naive method perform better than the forecast values under consideration.  \\   \hline 
		MBE  & \citet{van2017review} & The Mean Bias Error assess the average bias, where a large and positive MBE represents a large overestimate. However, it is scale dependent and lacks of information about the distribution of the errors. \\ \hline 
		MSE, RMSE & \citet{Willmott2005} \citet{van2017review} & The Mean Square Error and the Root Mean Square Error are scale dependent and more sensitive to outliers than MAE due to the squared error. They provide a quick insight into the variance and standard deviation of the errors.  \\ \hline 
		NRMSE & \citet{van2017review} & The Normalized Root Mean Square Error is a normalized RMSE. Common choices for the normalization factor are the mean or the range (the maximum minus the minimum) of the data. NRMSE is scale invariant but is more sensitive to outliers than MAPE due to the squared error. 
	\end{tabular}
\end{table*}

\begin{table*}[!htb]
	\renewcommand\thetable{11.2}
	\renewcommand{\arraystretch}{1.2}
	\caption{Probabilistic forecasting error metrics. } 
	\label{tab:EMproba}
	\begin{tabular}{m{1.1cm}||m{3cm}m{10.5cm}}
		EM   & Author &  Principles    \\  \hline  \hline
		CRPS   & \citet{van2017review} \citet{gneiting2014probabilistic} &  The Continuous Rank Probability Score measures both reliability and sharpness. It is the absolute error if the forecast is deterministic and therefore allows for comparison between probabilistic and point forecasts. \\ \hline 
		KSS   &  \citet{Hong2016}  & The Kolmogorov-Smirnov Statistic measures the unconditional coverage. The smallest KSS  indicates the best forecasted distribution. However, it is not very sensitive for establishing the distance between two distributions and does not evaluate forecasts sharpness or resolution. \\  \hline 
		PICP  &  \citet{van2017review} & The Prediction Interval Coverage Probability is a quantitative expression of reliability and should be higher than the nominal confidence level. The evaluation of the PICP alone is misleading, since a forecast with very wide prediction interval can result in a high PICP. \\ \hline 
		PINAW & \citet{van2017review} & The Prediction Interval Normalized Average Width quantitatively assesses the width of the prediction intervals. Usually high PINAW implies high PICP.   \\ \hline 
		PLF & \citet{van2017review} \citet{Hong2016} &  The Pinball Loss Function takes both reliability and sharpness into consideration and is specifically designed for quantile forecasts. A lower score indicates a better prediction interval.  \\ \hline 
		WS  &  \citet{Hong2016} & The Winkler Score allows a joint assessment of the unconditional coverage and interval width. It gives a penalty if an observation lies outside the constructed interval and rewards forecasts with a narrow prediction interval.
	\end{tabular}
\end{table*}

\end{document}